\documentclass[11pt]{article}
\usepackage[table]{xcolor}
\usepackage{multirow}
\usepackage[final]{acl}

\usepackage{times}
\usepackage{latexsym}

\usepackage{algorithm}
\usepackage{algpseudocode}
\usepackage{booktabs} 
\usepackage{amsmath}
\usepackage{float}
\usepackage[T1]{fontenc}

\usepackage[utf8]{inputenc}

\usepackage{microtype}

\usepackage{inconsolata}

\usepackage{graphicx}

\title{DuQUAD: A Dual-View Framework with Quality-Aware Evidence Pruning for Multi-Document Question Answering}

\author{
  \textbf{Yeonjun Choi}\textsuperscript{1}
  \thanks{Equal contribution.},
  \textbf{Seyeon Jeong}\textsuperscript{1}\footnotemark[1],
  \textbf{Jong Wook Kim}\textsuperscript{2}
  \thanks{Co-corresponding authors.},
  \textbf{Beakcheol Jang}\textsuperscript{1}\footnotemark[2]
\\
\\
  \textsuperscript{1}Graduate School of Information, Yonsei University, Seoul, South Korea
\\
  \textsuperscript{2}Department of Computer Science, Sangmyung University, Seoul, South Korea
\\
  \texttt{\{chlduswns99, seyeonjeong, bjang\}@yonsei.ac.kr}
  \quad
  \texttt{jkim@smu.ac.kr}
}

\begin{document}
\maketitle

\begin{abstract}
Large language models with retrieval-augmented generation systematically underutilize evidence in the middle of multi-document contexts, a phenomenon known as Lost in the Middle. Existing mitigation strategies, such as document reordering and attention steering, can reduce boundary bias but remain brittle in noisy settings, where inaccurate relevance estimation and spurious documents still obscure useful evidence in central positions. We propose DuQUAD, a multi-document QA framework that mitigates positional bias through dual-view reasoning and quality-aware evidence pruning. DuQUAD combines a Local Agent guided by a Structural Fusion Score with a Global Agent that reasons over the full document set, providing both targeted evidence recovery and broad contextual understanding. The framework further filters candidate evidence using explicit quality scoring and refines it at the sentence level to reduce noise. DuQUAD consistently outperforms strong Lost-in-the-Middle mitigation baselines, including hybrid retrieval and reranking, multi-agent reasoning, and context optimization methods, improving accuracy by 3.0\% on average over the strongest competing baseline and gold evidence recall by 9.4\%, while substantially reducing computational cost compared to existing multi-agent frameworks.
\end{abstract}

\section{Introduction}

Large Language Models (LLMs) have demonstrated strong performance on knowledge-intensive tasks \cite{achiam2023gpt, yang2025qwen3, grattafiori2024llama3herdmodels}, but remain limited by their static parametric knowledge, leading to issues such as obsolescence and hallucination. To address this, many approaches provide LLMs with external documents as additional context. However, even when relevant documents are available, LLMs often fail to use them effectively in multi-document settings due to a phenomenon known as Lost in the Middle (LiTM) \cite{liu2024lost}. LLMs exhibit strong positional bias, disproportionately attending to documents at the boundaries while underutilizing evidence in central positions, which degrades multi-document QA performance~\cite{liu2024lost, wang2024leave, hutter2025lost}. This degradation is further amplified by contextual noise. We use \textit{noise} to denote answer-irrelevant or weakly relevant documents in the retrieved set that disperse attention away from answer-bearing evidence and introduce distracting cues during reasoning.

\begin{figure}[t]
  \centering
  \includegraphics[width=0.99\columnwidth]{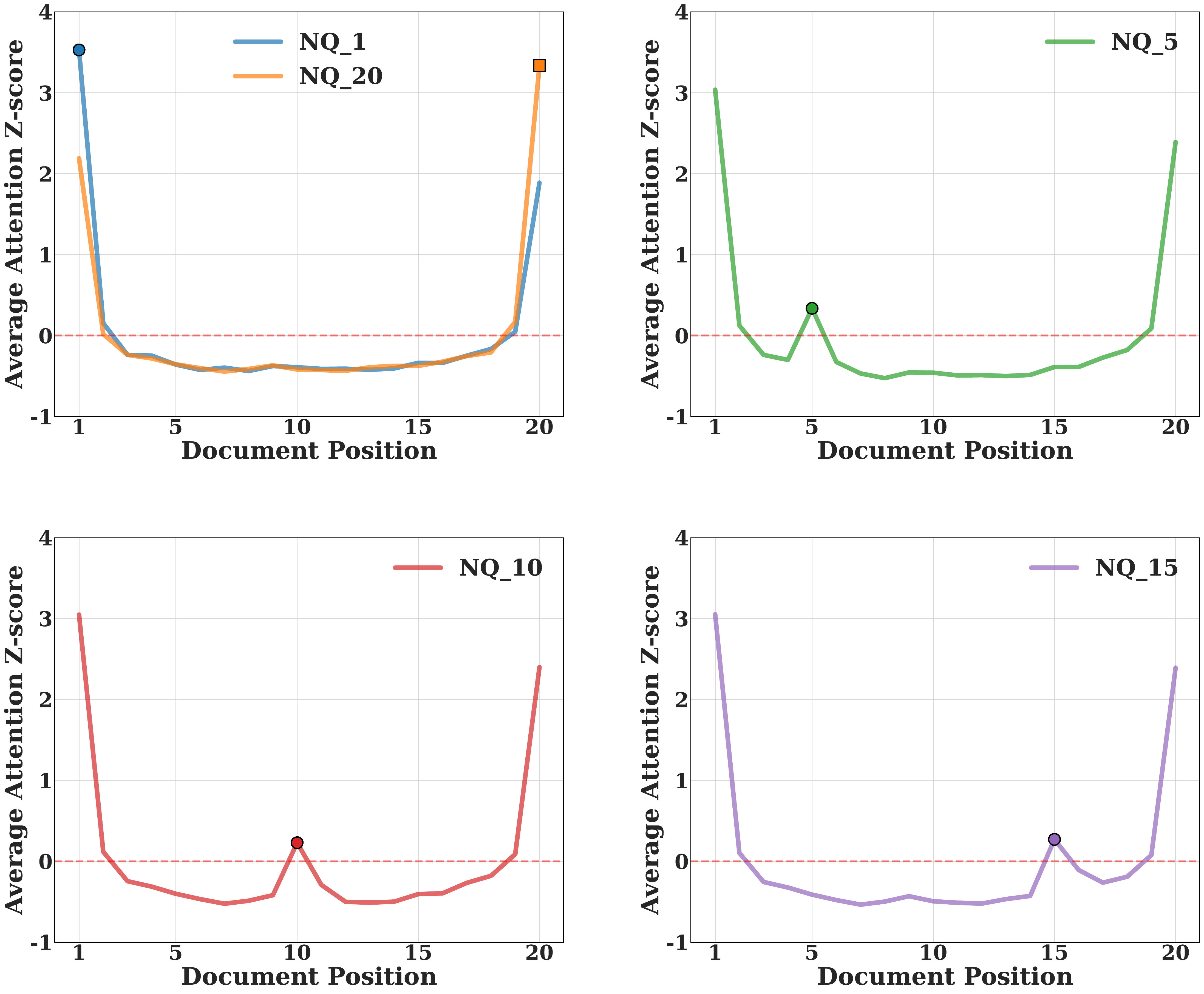}
  \caption{Attention z-score distribution across document positions on the Natural Questions (NQ) benchmark (NQ-1 to NQ-20, where NQ-$n$ denotes instances in which the gold document appears at the $n$-th position).
  }
  \label{fig:attention-wpattern}
\end{figure}

Consistent with the U-shaped positional bias \cite{hsieh2024found}, attention concentrates at the beginning and end of the context. When gold documents appear in intermediate positions, localized spikes form a W-shaped pattern (Figure~\ref{fig:attention-wpattern}), but boundary attention still dominates, leaving mid-context evidence underutilized.

To mitigate positional bias, most existing approaches rely on document reordering~\cite{peysakhovich2023attention, liu2024lost, jiang2024longllmlingua} or attention modification during training or inference \cite{hsieh2024found, jiang2024longllmlingua}. However, relying on weakly calibrated signals such as semantic similarity or raw attention renders these approaches vulnerable to noise, particularly when gold evidence is buried among weakly relevant distractors at mid-context positions.

Motivated by these localized spikes, we propose a framework that mitigates positional bias through dual-view reasoning and quality-aware pruning. We operationalize the W-shaped attention pattern as an attention contrast signal and integrate it with semantic and structural inference via a Structural Fusion Score (SFS) to prioritize gold documents. Our approach combines a Local Agent operating on prioritized documents with a Global Agent reasoning over the full context. Their outputs undergo DocScore-based filtering to prevent noise and echo chamber effects \cite{estornell2024multi}, before a Judge Agent synthesizes the final answer.

We evaluate our framework on multi-document QA benchmarks covering controlled position and multi-hop reasoning scenarios, comparing against recent baselines including hybrid retrieval and reranking, LiTM mitigation strategies, and multi-agent frameworks. Our contributions are summarized as follows:

\begin{itemize}
    \item We formalize the W-shaped attention pattern by adapting the edge-detection principle of the Laplacian operator \cite{marr1980theory} to capture sharp local attention spikes amidst smooth positional bias.

    \item We propose the Structural Fusion Score (SFS), a unified document scoring function that integrates attention contrast signals, semantic relevance, and structural inference, enabling identification and re-ranking of documents for local evidence selection.

    \item We introduce a Dual-View Agent Framework that mitigates positional bias by synergizing a global context view with a precise local view, integrated with quality-aware pruning to filter noise and prevent hallucination.

    \item Extensive experiments across three multi-document QA benchmarks show that DuQUAD improves accuracy by 3.0\% on average over the strongest competing baseline, achieves a 13.8\% gain over the stronger multi-agent baseline on NQ-15, and raises gold evidence recall by 9.4\%.
\end{itemize}



\begin{figure*}[t]
  \centering
  \includegraphics[width=0.99\linewidth]{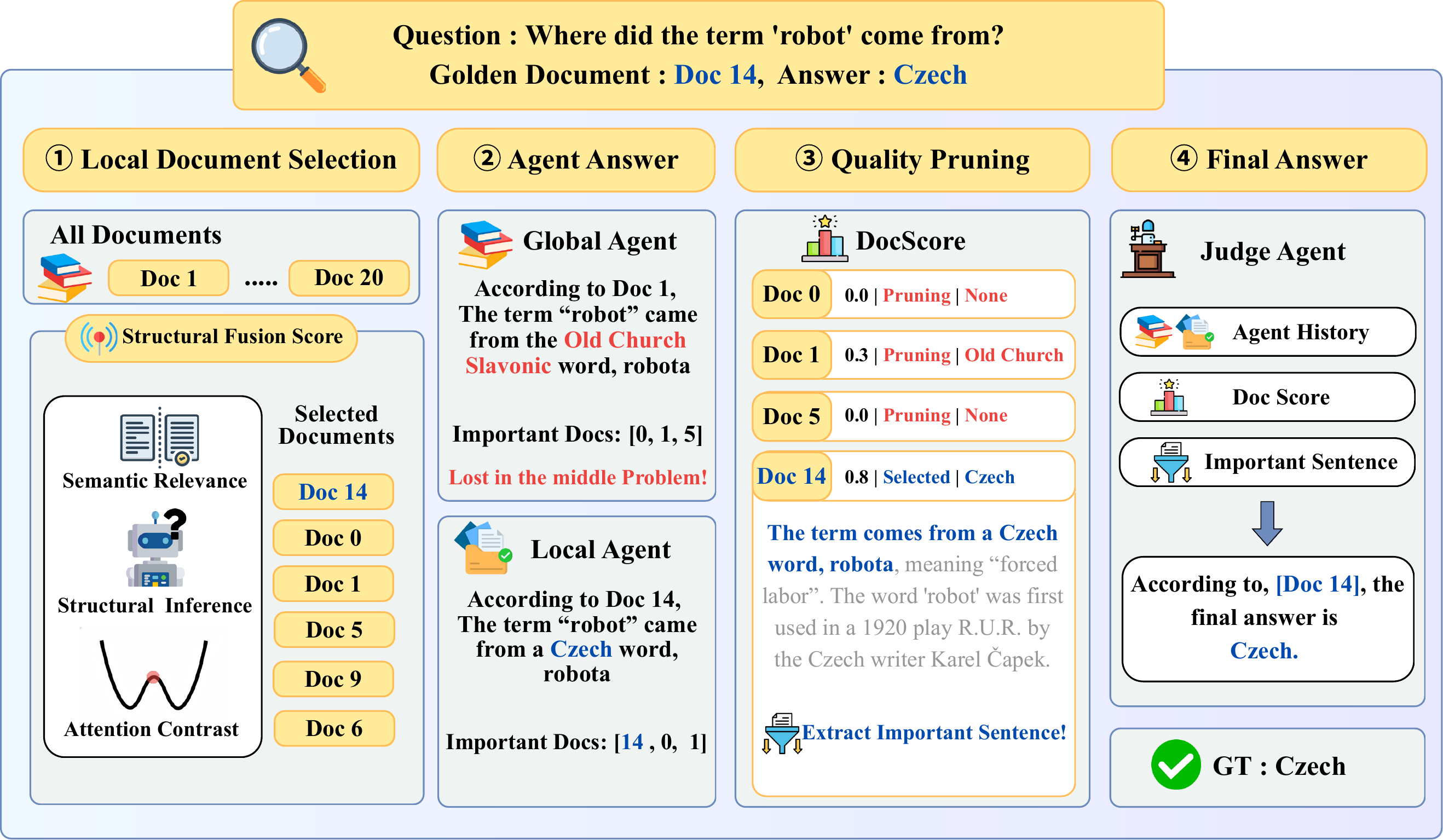}
  \caption{Overview of a dual-view multi-agent reasoning framework based on quality-aware evidence pruning. Given a question and a set of retrieved documents, the Structural Fusion Score (SFS) prioritizes a subset for the Local Agent, while the Global Agent operates on the full context. Selected documents are filtered via LLM-based DocScore, and key sentences are extracted. Finally, the Judge Agent aggregates reasoning histories and extracted evidence to produce the final answer.}
  \label{fig:fig2}
\end{figure*}

\section{Related Work}

\paragraph{Positional Bias in Large Language Models}
LLMs exhibit strong positional biases on multi-document inputs, tending to over-attend to tokens near the input boundaries while under-utilizing information in intermediate positions~\cite{liu2024lost, hutter2025lost}.
This phenomenon, known as Lost in the Middle (LiTM), is attributed to attention dynamics such as attention sinks and recency effects~\cite{ICLR2024_5e5fd18f, ICLR2025_f1b04fac}.
To mitigate these effects, prior work has proposed attention calibration and sorting techniques~\cite{peysakhovich2023attention, hsieh2024found}, as well as positional scaling and dimension-level bias correction~\cite{yu2024mitigate}.
However, these methods primarily target attention-level symptoms rather than how positional bias degrades evidence utilization in multi-document reasoning.

\paragraph{Training-based Approaches to LiTM Mitigation}
Training-based methods address positional bias through fine-tuning or additional supervision~\cite{liu2024lost, han2024lm}.
Position-agnostic training and multi-step QA supervision encourage models to prioritize content relevance over absolute position~\cite{he-etal-2024-never, pan2024not}, while full-context utilization training directly optimizes for mid-context evidence use~\cite{an2024make}.
Although effective, these approaches require substantial model adaptation and cannot be applied to off-the-shelf models at inference time.

\paragraph{Training-free Approaches to LiTM Mitigation}
Training-free methods mitigate positional bias at inference time without modifying model parameters~\cite{wang2024leave}.
Document reordering and reranking place relevant passages at salient positions prior to reasoning~\cite{peysakhovich2023attention, wang2025jina}, but do not address suppression during reasoning itself.
Context compression reduces input noise~\cite{jiang2024longllmlingua, verma2024contextual}, but risks discarding weak signals necessary for multi-hop reasoning.
These methods do not explicitly model how positional bias suppresses evidence during reasoning, leaving mid-context evidence vulnerable to being overlooked.

\paragraph{Multi-agent Reasoning for Multi-document QA}
Multi-agent reasoning extends single-agent LLMs by decomposing complex tasks into coordinated sub-problems, enabling role specialization and iterative refinement through agent collaboration. Prior work has explored various paradigms, including debate-based reasoning~\cite{du2024improving} and role-based decomposition (e.g., planner–solver–verifier)~\cite{chen-etal-2025-magicore}. Building on these ideas, structured approaches such as Tree of Agents~\cite{yu-etal-2025-tree} and MAIN-RAG~\cite{chang2025main} organize reasoning and retrieval into hierarchical or multi-stage pipelines. However, these methods primarily focus on task decomposition and document selection, and do not explicitly account for positional bias, leaving mid-context evidence still under-utilized.

\section{Methodology}

We address reasoning failures in multi-document QA: despite accurate retrieval, standard RAG models suffer from positional bias and the LiTM effect, where mid-context evidence is suppressed by noise. To reliably isolate gold evidence, we propose a hierarchical framework that integrates document prioritization, dual-view agent collaboration, and quality-aware evidence pruning, as illustrated in Figure~\ref{fig:fig2}.

\subsection{Structural Fusion Score (SFS)}

To identify gold documents under positional noise, we propose the Structural Fusion Score (SFS).
While attention over multi-document inputs typically exhibits a U-shaped bias, gold evidence forms a characteristic \emph{W-shaped} pattern with sharp local spikes \cite{hsieh2024found}.
SFS combines (1) semantic relevance, (2) structural inference, and (3) attention contrast.
The first two capture query--document alignment, while attention contrast emphasizes localized spikes robust to positional bias.

\subsubsection{Semantic Relevance ($\Phi_{\mathrm{sim}}$)}

To capture the semantic alignment between a query and documents, we use General Text Embeddings (GTE) \cite{li2023towards}\footnote{https://huggingface.co/Alibaba-NLP/gte-large-en-v1.5}, a strong open-source embedding model with stable semantic similarity performance. Given a query $Q$ and a document $D_i$, we compute their dense vector representations $\mathbf{v}_Q$ and $\mathbf{v}_{D_i}$ independently. The semantic relevance is quantified as their cosine similarity:

\begin{equation}
\Phi_{\mathrm{sim}, i}
=
\frac{\mathbf{v}_Q^\top \mathbf{v}_{D_i}}
{\|\mathbf{v}_Q\| \, \|\mathbf{v}_{D_i}\|}
\end{equation}

\subsubsection{Structural Inference \texorpdfstring{($\Phi_{\mathrm{ppl}}$)}{(Phi\_inf)}}

To capture structural inference signals, we evaluate the causal dependency of the query $Q=(q_1,\dots,q_{|Q|})$ on a document $D_i$ by measuring the likelihood of generating it conditioned on the document context. In all experiments, $\Phi_{\mathrm{ppl}}$ is computed using the same backbone LLM as the main reasoning model, which avoids introducing a separate scoring model and reduces cross-model inconsistency. This metric reflects intrinsic generative support beyond static embeddings.

\begin{equation}
\Phi_{\mathrm{ppl}, i}
=
-\frac{1}{|Q|}
\sum_{t=1}^{|Q|}
\log P(q_t \mid D_i, q_{<t})
\end{equation}

Here, $P(q_t \mid D_i, q_{<t})$ denotes the token-level generation probability assigned by the language model. Lower $\Phi_{\mathrm{ppl}}$ values indicate stronger contextual support for the query, serving as a robust indicator of gold evidence.

\subsubsection{Attention Contrast \texorpdfstring{($\Phi_{\mathrm{cont}}$)}{(Phi\_ac)}}

To isolate salient information peaks from positional noise, we compute a second-order local contrast over document-level attention scores. While absolute attention values across multiple documents are heavily entangled with positional bias, sharp local deviations more reliably indicate gold evidence, revealing characteristic W-shaped patterns.

$a_i$ is the average attention mass assigned from the final query token to the tokens of document $D_i$, aggregated over all layers and heads. Appendix~\ref{app:doc_attention} details this query-aware input construction and attention extraction procedure. Let $\mu_A$ and $\sigma_A$ be the mean and standard deviation of $a_i$ over the retrieved set $A=\{D_1,\dots,D_N\}$.
We first standardize the scores:

\[
z_i = \frac{a_i - \mu_A}{\sigma_A}.
\]

We then define the structural contrast as a second-order difference:
\[
\Phi_{\mathrm{cont}, i}
=
z_i - \frac{z_{i-1} + z_{i+1}}{2}.
\]
This formulation is analogous to a discrete Laplacian operator
\cite{gonzalez2009digital,marr1980theory}, suppressing smooth positional trends while emphasizing sharp local attention spikes.
At document boundaries where symmetric neighbors are not available, we adopt a one-sided finite difference scheme, detailed in Appendix~\ref{app:att_contrast}.

\subsubsection{Score Fusion ($\Phi$)}

We integrate heterogeneous signals using a weighted softmax-based fusion.
The final \textbf{Structural Fusion Score (SFS)} for document $D_i$ is defined as:

\begin{equation}
\Phi_i
=
\alpha \mathcal{S}(\Phi_{\mathrm{sim}, i})
+ \beta \mathcal{S}(-\Phi_{\mathrm{ppl}, i})
+ \gamma \mathcal{S}(\Phi_{\mathrm{cont}, i})
\end{equation}

where $\mathcal{S}(\cdot)$ denotes softmax applied over the retrieved documents.
We negate $\Phi_{\mathrm{ppl}}$ to ensure directional consistency.

Softmax addresses scale heterogeneity and induces distribution sharpening,
mapping disparate metrics into a unified probabilistic space while amplifying
high-confidence evidence.
We set $\alpha = 0.5$, $\beta = 1.0$, and $\gamma = 0.5$, placing greater
weight on structural inference signals. These weights are selected by grid
search on Natural Questions based on average gold evidence recall across
positions. As shown in Appendix~\ref{app:sfs_weight}, the chosen configuration
achieves the highest overall recall while remaining stable across nearby
settings.

\subsection{Dual-View Agent Collaboration}

Although SFS prioritizes relevant evidence, using only a filtered subset may lose global context and cross-document dependencies. To address this, we introduce a Dual-View Agent Collaboration framework that reasons from two complementary views.

The Global Agent reasons over the complete retrieved document set, while the Local Agent focuses on a compact, noise-reduced subset re-ranked by SFS. This design provides mutual robustness: the Local Agent recovers evidence suppressed by positional bias or noise, and the Global Agent compensates for information lost through filtering. Both agents generate cited answers independently, which are aggregated into a unified reasoning log for position-resilient multi-document reasoning. We use $k=6$ local documents based on a recall--accuracy trade-off analysis in Appendix~\ref{app:topk_analysis}.

\subsection{Quality-Aware Evidence Pruning}

While Dual-View reasoning maximizes recall, aggregating verbose reasoning logs and long-form documents can cause context overload, increasing the risk of hallucination. To address this issue, we introduce a two-stage quality-aware pruning strategy that separates evidence selection from quality verification.

\paragraph{Important Document Selection}
Each agent independently identifies a set of important documents based on its own reasoning process, directly from its generated output. Specifically, each agent returns up to three document IDs as important documents, together with its reasoning results, following a predefined JSON output format (Appendix~\ref{app:reproducibility}, Figure~\ref{fig:prompt1}).

Let $\mathcal{D}_{\mathrm{G}}$ and $\mathcal{D}_{\mathrm{L}}$ denote the sets of important documents
selected by the Global Agent and the Local Agent, respectively.
These sets are merged to form a candidate document pool as follows:
\begin{equation}
\mathcal{D}_{\mathrm{imp}} = \mathcal{D}_{\mathrm{G}} \cup \mathcal{D}_{\mathrm{L}}
\end{equation}

As a result, up to six candidate documents are obtained.
This stage serves as a coarse, recall-oriented filtering step that broadly preserves
answer-relevant evidence which might otherwise be missed due to positional bias
or contextual noise.

\paragraph{LLM-Based Quality Scoring}
Each candidate document $D_i \in \mathcal{D}_{\mathrm{imp}}$ is evaluated using
DocScore, an LLM-based document reliability score defined in this work, where
$p_i \in [0,1]$ estimates whether the question can be answered from the given
document alone. DocScore is produced by a dedicated Score Agent with a fixed
prompt, which returns a JSON-formatted confidence score together with an answer
span (Appendix~\ref{app:reproducibility}, Figure~\ref{fig:prompt3}).

We retain only high-confidence documents using a threshold-based rule:
\begin{equation}
\mathcal{D}_{\mathrm{keep}}
=
\{ D_i \in \mathcal{D}_{\mathrm{imp}} \mid p_i \ge \tau \} .
\end{equation}

We fix the threshold at $\tau = 0.4$ for all experiments. This value is selected
based on the threshold sensitivity analysis in Appendix~\ref{app:docscore_threshold}
(Table~\ref{tab:docscore_threshold_oprange}), where $\tau = 0.4$ is the highest threshold
that preserves recall while still enabling meaningful pruning. Lower thresholds admit
more irrelevant or weakly relevant documents, whereas higher thresholds risk
discarding gold evidence too early.

For each retained document, we generate a document-specific provisional answer and
select the top two sentences as supporting evidence based on word or keyword overlap
between the provisional answer and each sentence in the document. This sentence-level
evidence compression reduces the computational burden on the Judge Agent while
preserving the key information that directly supports final reasoning.

\subsection{Final Decision via Judge Agent}
The Judge Agent produces the final answer with a fixed prompt
(Appendix~\ref{app:reproducibility}, Figure~\ref{fig:prompt2}), jointly considering
(i) the reasoning histories of the Global and Local Agents,
(ii) document-level confidence scores (DocScore),
and (iii) sentence-level evidence spans extracted from retained documents.
Rather than relying on a single score or isolated evidence, the Judge integrates
these heterogeneous signals to make a consolidated decision.

Formally, given the question $Q$, the reasoning traces
$\mathcal{H}_{\mathrm{G}}$ and $\mathcal{H}_{\mathrm{L}}$,
the sentence-level evidence set $\mathcal{S}$,
and the associated DocScores $\mathbf{p}$ for documents in
$\mathcal{D}_{\mathrm{keep}}$, the final answer $\hat{y}$ is generated as:
\begin{equation}
\hat{y}
=
\mathrm{JudgeAgent}(
Q,
\mathcal{H}_{\mathrm{G}},
\mathcal{H}_{\mathrm{L}},
\mathcal{S},
\mathbf{p}
)
\end{equation}

By aggregating complementary agent outputs and evidence signals, the Judge Agent avoids over-reliance on any single source and supports robust answer generation.
The complete DuQUAD pipeline is summarized in Algorithm~\ref{alg:duquad}.

\begin{table*}[t]
\centering
\small
\renewcommand{\arraystretch}{1.3}
\setlength{\tabcolsep}{4pt}

\begin{tabular}{l|l|ccccc|ccc ccc}
\hline
\hline
\multirow{2}{*}{\textbf{Model}}
& \multicolumn{1}{l|}{\multirow{2}{*}{\textbf{Method}}}
& \textbf{NQ-1}
& \textbf{NQ-5}
& \textbf{NQ-10}
& \textbf{NQ-15}
& \textbf{NQ-20}
& \multicolumn{3}{c}{\textbf{MuSiQue}}
& \multicolumn{3}{c}{\textbf{HotpotQA}} \\
\cline{3-7} \cline{8-10} \cline{11-13}

& & Acc. & Acc. & Acc. & Acc. & Acc.
& Acc. & EM & F1
& Acc. & EM & F1 \\
\hline

\multirow{13}{*}{\shortstack[l]{\textbf{Llama-3.1-8B} \\ \textbf{-Instruct}}}

& Vanilla LLM
& 73.5 & 70.5 & 65.6 & 67.0 & 64.0
& 33.1 & \underline{19.9} & 27.3
& 73.8 & 52.4 & 68.4 \\

& SBERT
& 70.5 & 70.3 & 70.4 & 70.1 & 70.3
& 24.9 & 18.7 & 27.0
& 76.8 & 52.2 & 68.0 \\

& BM25
& 70.4 & 69.9 & 69.8 & 69.2 & 69.7
& 33.6 & 17.7 & 26.5
& 75.3 & 50.8 & 67.8 \\

& Jina-Reranker
& 69.0 & 68.4 & 69.2 & 69.4 & 68.6
& 29.3 & 17.6 & 27.0
& \underline{77.3} & 51.6 & 67.4 \\

& Qwen3-Emb.
& 69.3 & 69.3 & 69.3 & 69.3 & 69.3
& 34.7 & 18.0 & 27.0
& 76.8 & 50.3 & 67.2 \\

& OpenAI
& 70.2 & 70.5 & 71.4 & 69.9 & 70.5
& 34.7 & 18.3 & \underline{27.7}
& 76.7 & 52.5 & \underline{68.9} \\

& Hybrid (RRF)
& 71.9 & \underline{72.0} & \underline{72.1} & \underline{72.2} & \underline{72.3}
& 28.0 & 16.9 & 25.2
& 77.1 & \underline{53.8} & 68.7 \\

& \hspace{6pt}+ Rerank
& 71.4 & 71.4 & 71.4 & 71.4 & 71.4
& 28.8 & 17.8 & 26.8
& 77.2 & \textbf{54.4} & 68.8 \\

\cline{2-13}

& ToA
& 58.0 & 53.5 & 56.1 & 51.0 & 52.2
& 21.5 & 14.0 & 22.2
& 57.3 & 44.2 & 58.0 \\

& MAIN-RAG
& 62.5 & 64.4 & 63.1 & 62.3 & 61.4
& 19.0 & 16.2 & 23.7
& 60.5 & 46.6 & 61.6 \\

\cline{2-13}

& FiM
& \underline{75.3} & 67.0 & 65.6 & 64.7 & 65.4
& \underline{38.6} & 12.0 & 22.4
& 76.9 & 37.0 & 60.6 \\

& LongLLMLingua
& 67.4 & 68.9 & 67.1 & 67.9 & 66.4
& 30.8 & 15.7 & 24.9
& 70.2 & 49.7 & 65.6 \\

\cline{2-13}

& \cellcolor[rgb]{0.94, 0.97, 1.0}\textbf{DuQUAD}
& \cellcolor[rgb]{0.94, 0.97, 1.0}\textbf{77.3}
& \cellcolor[rgb]{0.94, 0.97, 1.0}\textbf{76.0}
& \cellcolor[rgb]{0.94, 0.97, 1.0}\textbf{75.2}
& \cellcolor[rgb]{0.94, 0.97, 1.0}\textbf{76.1}
& \cellcolor[rgb]{0.94, 0.97, 1.0}\textbf{76.7}
& \cellcolor[rgb]{0.94, 0.97, 1.0}\textbf{41.1}
& \cellcolor[rgb]{0.94, 0.97, 1.0}\textbf{20.6}
& \cellcolor[rgb]{0.94, 0.97, 1.0}\textbf{30.6}
& \cellcolor[rgb]{0.94, 0.97, 1.0}\textbf{80.2}
& \cellcolor[rgb]{0.94, 0.97, 1.0}52.6
& \cellcolor[rgb]{0.94, 0.97, 1.0}\textbf{70.2} \\

\hline
\hline

\multirow{13}{*}{\shortstack[l]{\textbf{Qwen-2.5-7B} \\ \textbf{-Instruct}}}

& Vanilla LLM
& 68.1 & 67.4 & 65.9 & 66.4 & 66.9
& 32.2 & 17.7 & 26.2
& 79.2 & 53.5 & 66.8 \\

& SBERT
& 66.5 & 66.4 & 66.1 & 66.6 & 66.2
& 29.6 & \underline{21.3} & \underline{29.0}
& 79.2 & \underline{57.5} & \underline{71.1} \\

& BM25
& 63.3 & 63.5 & 63.8 & 63.3 & 63.6
& 29.2 & 18.6 & 26.6
& 80.1 & 56.8 & 69.5 \\

& Jina-Reranker
& 69.0 & 68.4 & 68.1 & 68.1 & 68.4
& 29.4 & 20.2 & 28.1
& 78.2 & 56.7 & 70.2 \\

& Qwen3-Emb.
& 65.2 & 65.2 & 65.2 & 65.2 & 65.2
& 27.7 & 19.5 & 27.1
& 79.3 & 56.3 & 69.9 \\

& OpenAI
& 65.4 & 65.4 & 65.9 & 65.0 & 65.2
& 30.2 & 20.0 & 27.5
& \underline{80.3} & 57.2 & \underline{71.1} \\

& Hybrid (RRF)
& 67.5 & 67.5 & 67.4 & 67.5 & 67.7
& 28.4 & 19.5 & 27.3
& 79.0 & 55.1 & 68.4 \\

& \hspace{4pt}+ Rerank
& 69.9 & \underline{69.9} & \underline{69.9} & \underline{69.9} & \underline{69.9}
& 29.9 & 19.9 & 27.7
& 77.8 & 56.9 & 70.2 \\

\cline{2-13}

& ToA
& 66.0 & 51.0 & 46.1 & 42.5 & 44.5
& 13.1 & 13.5 & 21.3
& 50.1 & 40.0 & 51.6 \\

& MAIN-RAG
& 62.0 & 60.8 & 60.1 & 61.5 & 61.9
& 27.0 & 14.7 & 22.3
& 68.0 & 52.4 & 65.3 \\

\cline{2-13}

& FiM
& \underline{73.6} & 61.0 & 59.0 & 58.5 & 58.7
& \underline{32.4} & 12.6 & 21.3
& 74.5 & 36.2 & 54.2 \\

& LongLLMLingua
& 63.8 & 64.9 & 65.2 & 64.7 & 65.1
& 25.6 & 15.7 & 24.9
& 67.6 & 51.5 & 61.9 \\

\cline{2-13}

& \cellcolor[rgb]{0.94, 0.97, 1.0}\textbf{DuQUAD}
& \cellcolor[rgb]{0.94, 0.97, 1.0}\textbf{75.5}
& \cellcolor[rgb]{0.94, 0.97, 1.0}\textbf{72.6}
& \cellcolor[rgb]{0.94, 0.97, 1.0}\textbf{72.6}
& \cellcolor[rgb]{0.94, 0.97, 1.0}\textbf{70.2}
& \cellcolor[rgb]{0.94, 0.97, 1.0}\textbf{75.1}
& \cellcolor[rgb]{0.94, 0.97, 1.0}\textbf{36.7}
& \cellcolor[rgb]{0.94, 0.97, 1.0}\textbf{22.4}
& \cellcolor[rgb]{0.94, 0.97, 1.0}\textbf{31.0}
& \cellcolor[rgb]{0.94, 0.97, 1.0}\textbf{82.0}
& \cellcolor[rgb]{0.94, 0.97, 1.0}\textbf{60.7}
& \cellcolor[rgb]{0.94, 0.97, 1.0}\textbf{76.0} \\

\hline
\hline
\end{tabular}

\caption{
Main experimental results comparing Llama-3.1-8B-Instruct and Qwen2.5-7B-Instruct.
The highest score in each column is shown in \textbf{bold}, and the second-highest score is \underline{underlined}.
}
\label{tab:main_results}
\end{table*}

\begin{figure*}[t]
  \centering
  \includegraphics[width=0.95\linewidth]{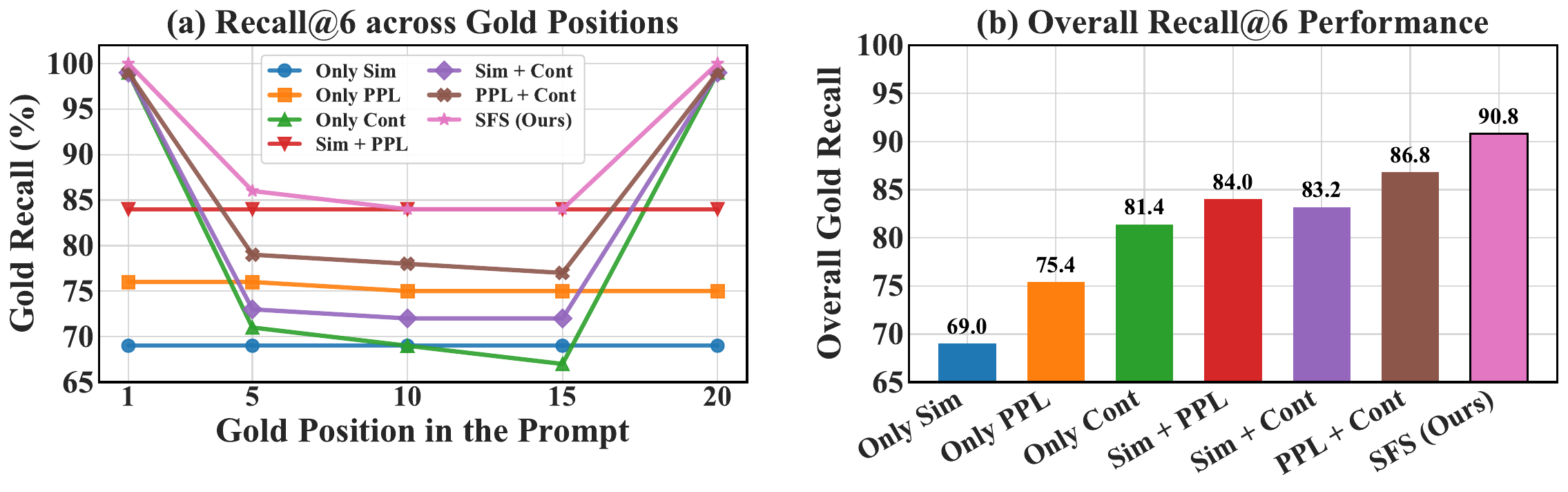}
\caption{Impact of Score Fusion on Robustness. Sim, PPL, and Cont denote the three SFS signals: Semantic Relevance, Structural Inference, and Attention Contrast, respectively. (a) Recall of the gold document by gold position under the Top-6 setting. (b) Overall recall aggregated across all gold positions in (a). Computed on 100 NQ examples per gold position.}
    \label{fig:fig3}
\end{figure*}

\section{Experiments}

\subsection{Experimental Setup}

\paragraph{Backbone Model}
We use Llama-3.1-8B-Instruct \cite{grattafiori2024llama3herdmodels}\footnote{\url{https://huggingface.co/meta-llama/Llama-3.1-8B-Instruct}} and Qwen2.5-7B-Instruct \cite{qwen2.5}\footnote{\url{https://huggingface.co/Qwen/Qwen2.5-7B-Instruct}} as backbone models. Both provide sufficient context windows and are applied consistently across all baselines to ensure fair comparison. Main results are reported for both models, while additional analyses use Llama-3.1-8B-Instruct by default.

\paragraph{Datasets}
We evaluate DuQUAD on three multi-document QA benchmarks. For Natural Questions (NQ) \cite{kwiatkowski-etal-2019-natural}, we follow the controlled positional bias protocol of \cite{liu2024lost}, placing one gold document at position $n \in \{1, 5, 10, 15, 20\}$ among 19 distractors. MuSiQue \cite{trivedi2022musique} requires connected reasoning across 2--4 hops, where each question provides 2--4 supporting paragraphs among distractors and no single paragraph suffices to answer independently. HotpotQA \cite{yang2018hotpotqa} requires bridge or comparison reasoning, evaluated in the distractor setting with two gold paragraphs among eight distractors. Detailed dataset characteristics, including input context statistics, are provided in Appendix~\ref{app:dataset} (Table~\ref{tab:dataset_statistics}), while per-experiment sample sizes are summarized in Table~\ref{tab:experiment_settings}.

\paragraph{Evaluation Metrics}
Following Lost in the Middle \cite{liu2024lost}, we evaluate performance using accuracy, defined as whether the generated response contains the ground-truth answer string. This metric is commonly used for multi-document QA with free-form outputs, focusing on answer presence rather than exact matching. For multi-hop reasoning tasks, we additionally report Exact Match (EM) for strict identity and F1 score for word overlap to assess generation precision.

\subsection{Baselines}

We compare our approach against baselines across three categories:
(1) reordering and reranking methods including SBERT \cite{reimers-gurevych-2019-sentence},
BM25 \cite{crestani1998document}, OpenAI embeddings \cite{openai2024textembed3},
Jina-Reranker \cite{wang2025jina}, Qwen3-Embedding \cite{zhang2025qwen3},
and two hybrid variants: Hybrid (RRF), which combines BM25 and Qwen3-Embedding
rankings using Reciprocal Rank Fusion (RRF) \cite{cormack2009reciprocal},
and Hybrid (RRF) + Rerank, which further applies Jina-Reranker to the fused ranking;
(2) multi-agent reasoning frameworks such as Tree of Agents (ToA) \cite{yu-etal-2025-tree}
and MAIN-RAG \cite{chang2025main};
and (3) context optimization techniques such as Found in the Middle (FiM)
\cite{hsieh2024found} and LongLLMLingua \cite{jiang2024longllmlingua}.
Implementation details are provided in Appendix~\ref{app:baseline}.

\begin{figure}[t]
  \centering
  \includegraphics[width=\linewidth]{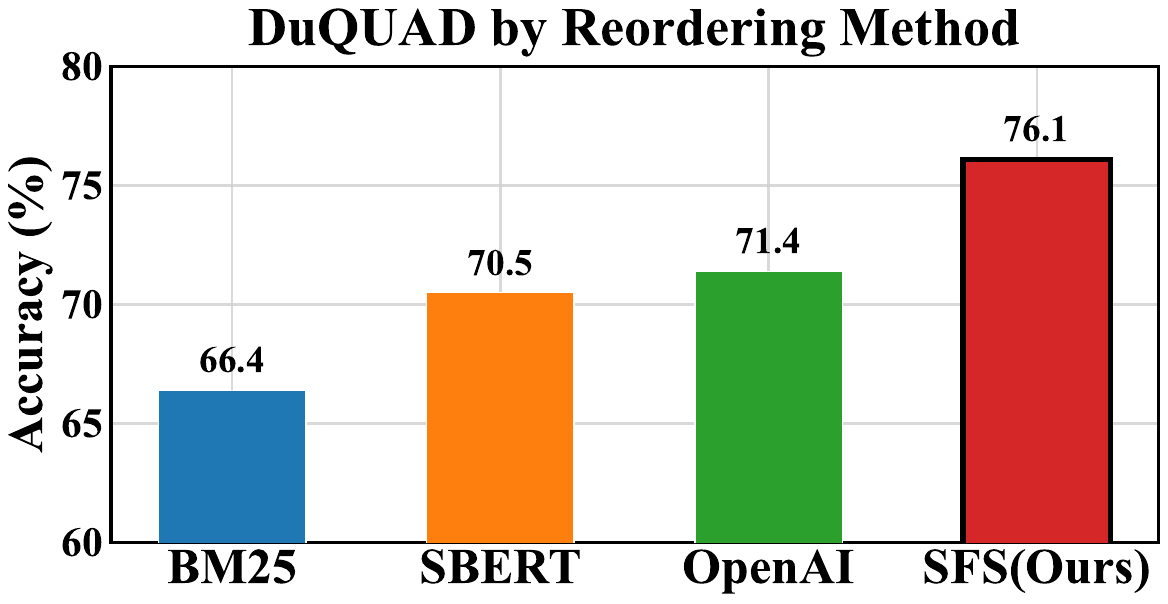}
\caption{
Impact of representative local document selection scores within DuQUAD on NQ-15.
We compare sparse and dense retrieval-based alternatives against the proposed
Structural Fusion Score (SFS) in terms of answer accuracy.
}
  \label{fig:fig5}
  
\end{figure}

\section{Results}
\subsection{Overall Performance}
Table~\ref{tab:main_results} reports accuracy on NQ, MuSiQue, and HotpotQA,
with EM and F1 scores additionally reported for multi-hop tasks.

On NQ, the Vanilla LLM shows strong positional sensitivity, with accuracy
varying substantially across gold-document positions. Reordering and reranking
baselines reduce this positional variation, with the hybrid variants showing
particularly stable performance across positions. Nevertheless, DuQUAD
consistently achieves the highest accuracy across all NQ settings under both
backbones. Context optimization and multi-agent baselines also remain
challenged by LiTM: FiM degrades substantially at intermediate positions,
while ToA and MAIN-RAG remain below DuQUAD despite their more complex
reasoning pipelines. These results suggest that stronger document ranking,
repositioning, or reasoning decomposition alone does not fully address
positional underutilization in multi-document reasoning.

Overall, DuQUAD achieves the best accuracy across all 14 accuracy settings
in Table~\ref{tab:main_results}. It improves over the strongest competing
baseline in each setting by 3.0 percentage points on average. The same trend
holds for both Llama-3.1-8B-Instruct and Qwen2.5-7B-Instruct, suggesting that
the gains are not specific to a single backbone.

On MuSiQue and HotpotQA, DuQUAD also achieves the best accuracy under both
backbones. With Llama-3.1-8B-Instruct, it improves over the strongest
competing baseline by 2.5 percentage points on MuSiQue and 2.9 percentage
points on HotpotQA. DuQUAD further achieves the highest F1 on both multi-hop
benchmarks. The only metric-level exception under the Llama backbone is
HotpotQA EM, where Hybrid (RRF) + Rerank achieves 54.4\%, compared with
52.6\% for DuQUAD. This narrow EM advantage is consistent with HotpotQA's
relatively compact evidence setting, where strong document ordering can place
the two gold paragraphs at salient positions, while strict exact-match scoring
may penalize answers that are semantically correct but differ in surface form.
With Qwen2.5-7B-Instruct, DuQUAD achieves the highest accuracy, EM, and F1
on both multi-hop benchmarks. These results demonstrate that DuQUAD maintains
robust performance across both controlled positional settings and multi-hop
reasoning tasks. Paired bootstrap resampling further shows that DuQUAD's gains
are statistically significant in 35 out of 36 comparisons ($p<0.05$), with
detailed results in Appendix~\ref{statistic}.

\subsection{Gold Evidence Recall and DocScore Stability}
We evaluate local document selection using gold evidence recall,
defined as whether the gold document is included in the candidate set selected
for the Local Agent. As shown in Figure~\ref{fig:fig3}, Attention Contrast
achieves the highest recall among single-signal approaches at 81.4\%,
surpassing Semantic Relevance with 69.0\% and Structural Inference with 75.4\%.
In contrast, the proposed Structural Fusion Score (SFS) attains the highest
overall recall of 90.8\%, yielding a 4.0\% improvement over the strongest
multi-signal baseline that combines Structural Inference and Attention Contrast,
and a 9.4\% gain over the best single-signal method. These results indicate that
integrating complementary signals enables more robust identification of gold
evidence across different document positions.

DocScore further provides a stable quality signal for evidence pruning.
Across NQ-1/5/10/15/20, gold documents consistently receive higher DocScores
than non-gold documents and are selected as Important Documents more frequently
under both Llama and Qwen backbones. Cross-model analysis further shows 90.4\%
agreement on gold evidence selection, suggesting that DocScore captures a stable
answerability signal rather than backbone-specific behavior. Detailed statistics
are provided in Appendix~\ref{app:cross_model}.

\begin{table}[t]
    \centering
    \resizebox{\columnwidth}{!}{%
        \begin{tabular}{l|ccc}
            \hline
            Method & NQ-15 & MuSiQue & HotpotQA \\
            \hline
            w/o Local Agent & 69.8 & 38.1 & 79.1 \\
            w/o Global Agent & 71.1 & 36.1 & 80.0 \\
            w/o Quality Pruning & 71.3 & 34.6 & 79.8 \\
            
            \textbf{DuQUAD} & \textbf{76.1} & \textbf{41.1} & \textbf{80.2} \\
            \hline
        \end{tabular}%
    }
    \caption{Effect of the Local Agent, Global Agent, and Quality-Aware Pruning. Ablation analysis across three datasets evaluating the impact on accuracy when omitting each component compared to the complete DuQUAD method. }
    \label{tab:localquality}
\end{table}

\begin{figure}[t]
  \centering
  \includegraphics[width=\linewidth]{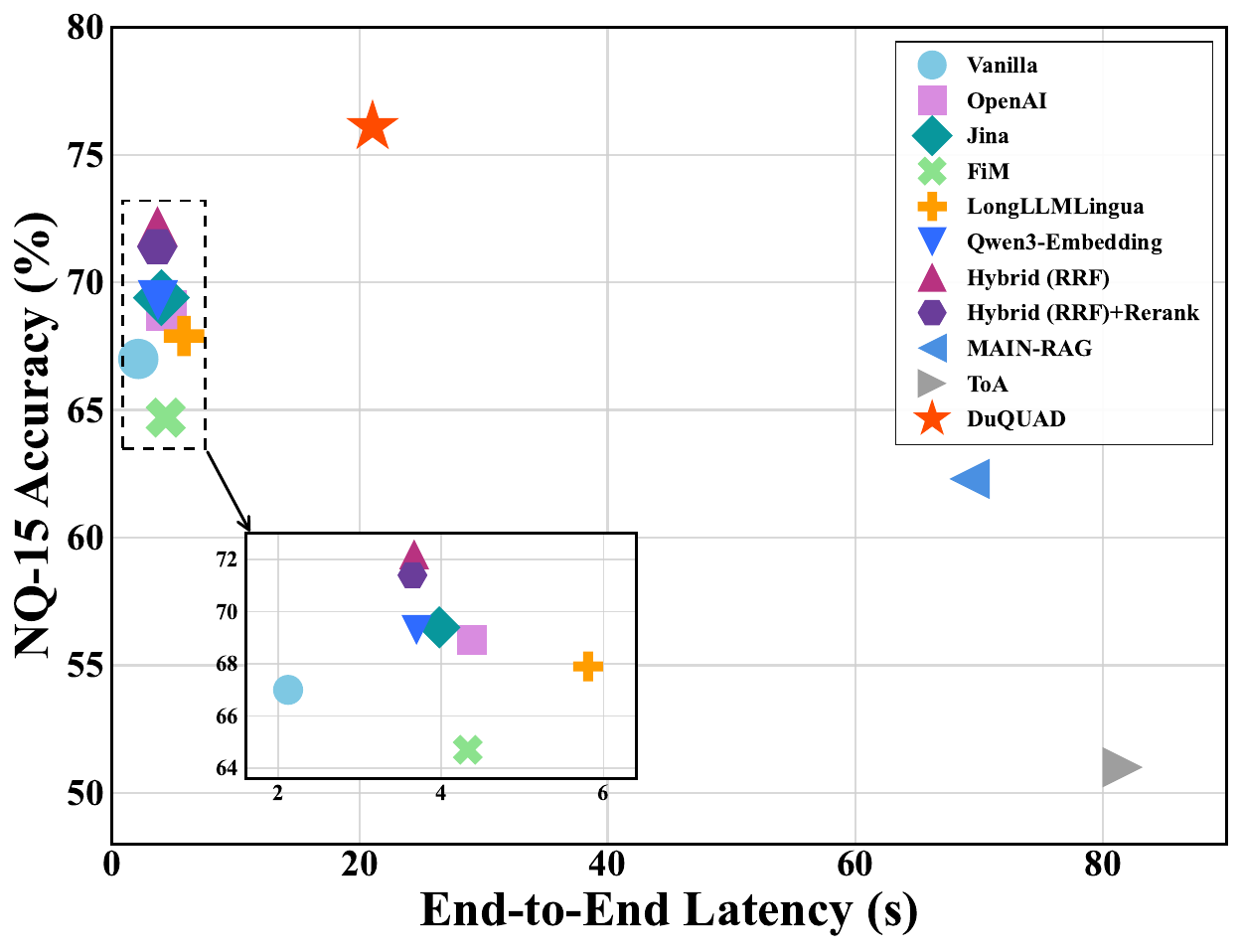}
 \caption{
Accuracy--latency trade-off of representative baselines on NQ-15.
The inset magnifies the low-latency region for clearer comparison among
retrieval, reranking, and context optimization methods.
}
  \label{fig:cost}
\end{figure}

\subsection{Ablation Analysis}

Table~\ref{tab:localquality} reports ablation results evaluating the contributions
of the Local Agent, the Global Agent, and quality-aware evidence pruning.
Removing any component consistently degrades performance, confirming that DuQUAD's
gains come from the interaction of complementary reasoning views and explicit
evidence filtering. The largest drop on NQ-15 occurs when the Local Agent is removed,
indicating that global context reasoning alone is insufficient when answer-bearing
evidence is embedded among many distractors. In contrast, MuSiQue shows larger
drops when the Global Agent or quality-aware pruning is removed, suggesting that
multi-hop reasoning requires both cross-document context preservation and filtering
of noisy or weakly aligned reasoning traces. The smaller changes on HotpotQA suggest
less headroom under its smaller candidate set and more explicit supporting facts.

Figure~\ref{fig:fig5} further analyzes document prioritization under the NQ-15 setting
using representative local document selection scores.
Among the evaluated alternatives, OpenAI achieves the highest accuracy
at 71.4\%, compared with 76.1\% for the proposed Structural Fusion Score (SFS).
These results show that replacing SFS with standard sparse or dense retrieval scores
reduces performance, supporting the benefit of combining semantic relevance,
structural inference, and attention contrast for local document prioritization.
More broadly, DuQUAD integrates this SFS-based local prioritization with dual-view
reasoning and quality-aware pruning.
Appendix~\ref{app:failure} provides quantitative disagreement analysis and qualitative
examples of Local--Global complementarity and joint failure cases.


\subsection{Cost Efficiency Analysis}

We analyze the cost-efficiency of DuQUAD with respect to end-to-end latency on NQ-15.
Figure~\ref{fig:cost} shows the accuracy--latency trade-off across representative baselines.
Among lightweight retrieval and reranking methods, Hybrid (RRF) provides the strongest accuracy--latency trade-off, achieving 72.2\% accuracy with an end-to-end latency of 3.67\,s.
DuQUAD further improves accuracy to 76.1\%, at the cost of additional inference latency.

Compared with existing multi-agent frameworks, however, DuQUAD is both more accurate and substantially more efficient:
it improves accuracy by 13.8\% over MAIN-RAG and 25.1\% over ToA while reducing end-to-end latency by 3.3$\times$ and 3.9$\times$, respectively.
These results show that DuQUAD incurs additional overhead relative to lightweight retrieval and reranking methods, but achieves higher accuracy while remaining substantially more efficient than existing multi-agent frameworks.
As the Local and Global Agents operate independently, they can also be executed in parallel to further reduce latency.
Detailed cost statistics are provided in Appendix~\ref{app:cost}.

\section{Conclusion}
We investigated the limitations of multi-document QA under positional bias, specifically the Lost in the Middle phenomenon. To address it, we propose DuQUAD, a dual-view reasoning framework that pairs a Local Agent operating on a prioritized evidence subset with a Global Agent that reasons over the full document set. Local selection is guided by the Structural Fusion Score, which integrates semantic relevance, structural inference, and attention contrast to robustly identify gold documents despite positional noise. Quality-aware pruning further removes low-confidence evidence before final reasoning, and together these components raise gold evidence recall to 90.8\%. As a result, DuQUAD mitigates positional bias and achieves the best accuracy among the evaluated baselines across all benchmarks, while requiring substantially less computation than existing multi-agent frameworks. These results demonstrate that DuQUAD offers a favorable balance between accuracy and cost for robust reasoning in multi-document QA under positional bias.

\section{Limitations}

While DuQUAD effectively mitigates positional bias in multi-document QA settings, it also has several limitations.
First, DuQUAD employs an explicit evidence pruning mechanism to suppress
noisy or low-confidence documents.
Although this improves robustness against positional noise,
it may reduce contextual coverage when relevant evidence is weakly expressed
or distributed across multiple documents, particularly for complex
multi-hop questions.

Second, DuQUAD relies on the availability of answer-bearing documents
within the initial retrieved set.
Similar to other retrieval-augmented frameworks, its performance is bounded
by retrieval quality: if relevant documents are not retrieved,
the framework cannot recover missing evidence through internal reasoning alone.

Third, even when answer-bearing evidence is correctly retrieved and scored,
the final synthesis stage may still fail in complex multi-hop questions
where the reasoning chain must track closely related intermediate entities,
as reflected in the quantitative and qualitative analyses in
Appendix~\ref{app:failure}.

Finally, our evaluation is limited to Llama-3.1-8B-Instruct and
Qwen2.5-7B-Instruct; broader evaluation across model families, capabilities,
and sizes would further assess the generalizability of DuQUAD.

These limitations reflect inherent trade-offs in multi-document QA systems
that balance contextual coverage, noise suppression, and reasoning stability.

\section*{Acknowledgments}

This work was supported by the National Research Foundation of Korea (NRF) under Grant No. RS-2023-00273751 and by the Institute of Information \& Communications Technology Planning \& Evaluation (IITP) under Grant No. RS-2024-00397085, funded by the Korean government.

\bibliography{custom}

@article{achiam2023gpt,
  title={Gpt-4 technical report},
  author={Achiam, Josh and Adler, Steven and Agarwal, Sandhini and Ahmad, Lama and Akkaya, Ilge and Aleman, Florencia Leoni and Almeida, Diogo and Altenschmidt, Janko and Altman, Sam and Anadkat, Shyamal and others},
  journal={arXiv preprint arXiv:2303.08774},
  year={2023}
}

@inproceedings{hsieh2024found,
  title={Found in the middle: Calibrating positional attention bias improves long context utilization},
  author={Hsieh, Cheng-Yu and Chuang, Yung-Sung and Li, Chun-Liang and Wang, Zifeng and Le, Long and Kumar, Abhishek and Glass, James and Ratner, Alexander and Lee, Chen-Yu and Krishna, Ranjay and others},
  booktitle={Findings of the Association for Computational Linguistics: ACL 2024},
  pages={14982--14995},
  year={2024}
}

@article{peysakhovich2023attention,
  title={Attention sorting combats recency bias in long context language models},
  author={Peysakhovich, Alexander and Lerer, Adam},
  journal={arXiv preprint arXiv:2310.01427},
  year={2023}
}

@article{liu2024lost,
  title={Lost in the middle: How language models use long contexts},
  author={Liu, Nelson F and Lin, Kevin and Hewitt, John and Paranjape, Ashwin and Bevilacqua, Michele and Petroni, Fabio and Liang, Percy},
  journal={Transactions of the Association for Computational Linguistics},
  volume={12},
  pages={157--173},
  year={2024}
}

@inproceedings{jiang2024longllmlingua,
  title={Longllmlingua: Accelerating and enhancing llms in long context scenarios via prompt compression},
  author={Jiang, Huiqiang and Wu, Qianhui and Luo, Xufang and Li, Dongsheng and Lin, Chin-Yew and Yang, Yuqing and Qiu, Lili},
  booktitle={Proceedings of the 62nd Annual Meeting of the Association for Computational Linguistics (Volume 1: Long Papers)},
  pages={1658--1677},
  year={2024}
}

@inproceedings{chang2025main,
  title={Main-rag: Multi-agent filtering retrieval-augmented generation},
  author={Chang, Chia-Yuan and Jiang, Zhimeng and Rakesh, Vineeth and Pan, Menghai and Yeh, Chin-Chia Michael and Wang, Guanchu and Hu, Mingzhi and Xu, Zhichao and Zheng, Yan and Das, Mahashweta and others},
  booktitle={Proceedings of the 63rd Annual Meeting of the Association for Computational Linguistics (Volume 1: Long Papers)},
  pages={2607--2622},
  year={2025}
}

@article{yang2025qwen3,
  title={Qwen3 technical report},
  author={Yang, An and Li, Anfeng and Yang, Baosong and Zhang, Beichen and Hui, Binyuan and Zheng, Bo and Yu, Bowen and Gao, Chang and Huang, Chengen and Lv, Chenxu and others},
  journal={arXiv preprint arXiv:2505.09388},
  year={2025}
}

@misc{grattafiori2024llama3herdmodels,
      title={The Llama 3 Herd of Models}, 
      author={Aaron Grattafiori and Abhimanyu Dubey and Abhinav Jauhri and Abhinav Pandey and Abhishek Kadian and Ahmad Al-Dahle and Aiesha Letman and Akhil Mathur and Alan Schelten and Alex Vaughan and Amy Yang and Angela Fan and Anirudh Goyal and Anthony Hartshorn and Aobo Yang and Archi Mitra and Archie Sravankumar and Artem Korenev and Arthur Hinsvark and Arun Rao and Aston Zhang and Aurelien Rodriguez and Austen Gregerson and Ava Spataru and Baptiste Roziere and Bethany Biron and Binh Tang and Bobbie Chern and Charlotte Caucheteux and Chaya Nayak and Chloe Bi and Chris Marra and Chris McConnell and Christian Keller and Christophe Touret and Chunyang Wu and Corinne Wong and Cristian Canton Ferrer and Cyrus Nikolaidis and Damien Allonsius and Daniel Song and Danielle Pintz and Danny Livshits and Danny Wyatt and David Esiobu and Dhruv Choudhary and Dhruv Mahajan and Diego Garcia-Olano and Diego Perino and Dieuwke Hupkes and Egor Lakomkin and Ehab AlBadawy and Elina Lobanova and Emily Dinan and Eric Michael Smith and Filip Radenovic and Francisco Guzmán and Frank Zhang and Gabriel Synnaeve and Gabrielle Lee and Georgia Lewis Anderson and Govind Thattai and Graeme Nail and Gregoire Mialon and Guan Pang and Guillem Cucurell and Hailey Nguyen and Hannah Korevaar and Hu Xu and Hugo Touvron and Iliyan Zarov and Imanol Arrieta Ibarra and Isabel Kloumann and Ishan Misra and Ivan Evtimov and Jack Zhang and Jade Copet and Jaewon Lee and Jan Geffert and Jana Vranes and Jason Park and Jay Mahadeokar and Jeet Shah and Jelmer van der Linde and Jennifer Billock and Jenny Hong and Jenya Lee and Jeremy Fu and Jianfeng Chi and Jianyu Huang and Jiawen Liu and Jie Wang and Jiecao Yu and Joanna Bitton and Joe Spisak and Jongsoo Park and Joseph Rocca and Joshua Johnstun and Joshua Saxe and Junteng Jia and Kalyan Vasuden Alwala and Karthik Prasad and Kartikeya Upasani and Kate Plawiak and Ke Li and Kenneth Heafield and Kevin Stone and Khalid El-Arini and Krithika Iyer and Kshitiz Malik and Kuenley Chiu and Kunal Bhalla and Kushal Lakhotia and Lauren Rantala-Yeary and Laurens van der Maaten and Lawrence Chen and Liang Tan and Liz Jenkins and Louis Martin and Lovish Madaan and Lubo Malo and Lukas Blecher and Lukas Landzaat and Luke de Oliveira and Madeline Muzzi and Mahesh Pasupuleti and Mannat Singh and Manohar Paluri and Marcin Kardas and Maria Tsimpoukelli and Mathew Oldham and Mathieu Rita and Maya Pavlova and Melanie Kambadur and Mike Lewis and Min Si and Mitesh Kumar Singh and Mona Hassan and Naman Goyal and Narjes Torabi and Nikolay Bashlykov and Nikolay Bogoychev and Niladri Chatterji and Ning Zhang and Olivier Duchenne and Onur Çelebi and Patrick Alrassy and Pengchuan Zhang and Pengwei Li and Petar Vasic and Peter Weng and Prajjwal Bhargava and Pratik Dubal and Praveen Krishnan and Punit Singh Koura and Puxin Xu and Qing He and Qingxiao Dong and Ragavan Srinivasan and Raj Ganapathy and Ramon Calderer and Ricardo Silveira Cabral and Robert Stojnic and Roberta Raileanu and Rohan Maheswari and Rohit Girdhar and Rohit Patel and Romain Sauvestre and Ronnie Polidoro and Roshan Sumbaly and Ross Taylor and Ruan Silva and Rui Hou and Rui Wang and Saghar Hosseini and Sahana Chennabasappa and Sanjay Singh and Sean Bell and Seohyun Sonia Kim and Sergey Edunov and Shaoliang Nie and Sharan Narang and Sharath Raparthy and Sheng Shen and Shengye Wan and Shruti Bhosale and Shun Zhang and Simon Vandenhende and Soumya Batra and Spencer Whitman and Sten Sootla and Stephane Collot and Suchin Gururangan and Sydney Borodinsky and Tamar Herman and Tara Fowler and Tarek Sheasha and Thomas Georgiou and Thomas Scialom and Tobias Speckbacher and Todor Mihaylov and Tong Xiao and Ujjwal Karn and Vedanuj Goswami and Vibhor Gupta and Vignesh Ramanathan and Viktor Kerkez and Vincent Gonguet and Virginie Do and Vish Vogeti and Vítor Albiero and Vladan Petrovic and Weiwei Chu and Wenhan Xiong and Wenyin Fu and Whitney Meers and Xavier Martinet and Xiaodong Wang and Xiaofang Wang and Xiaoqing Ellen Tan and Xide Xia and Xinfeng Xie and Xuchao Jia and Xuewei Wang and Yaelle Goldschlag and Yashesh Gaur and Yasmine Babaei and Yi Wen and Yiwen Song and Yuchen Zhang and Yue Li and Yuning Mao and Zacharie Delpierre Coudert and Zheng Yan and Zhengxing Chen and Zoe Papakipos and Aaditya Singh and Aayushi Srivastava and Abha Jain and Adam Kelsey and Adam Shajnfeld and Adithya Gangidi and Adolfo Victoria and Ahuva Goldstand and Ajay Menon and Ajay Sharma and Alex Boesenberg and Alexei Baevski and Allie Feinstein and Amanda Kallet and Amit Sangani and Amos Teo and Anam Yunus and Andrei Lupu and Andres Alvarado and Andrew Caples and Andrew Gu and Andrew Ho and Andrew Poulton and Andrew Ryan and Ankit Ramchandani and Annie Dong and Annie Franco and Anuj Goyal and Aparajita Saraf and Arkabandhu Chowdhury and Ashley Gabriel and Ashwin Bharambe and Assaf Eisenman and Azadeh Yazdan and Beau James and Ben Maurer and Benjamin Leonhardi and Bernie Huang and Beth Loyd and Beto De Paola and Bhargavi Paranjape and Bing Liu and Bo Wu and Boyu Ni and Braden Hancock and Bram Wasti and Brandon Spence and Brani Stojkovic and Brian Gamido and Britt Montalvo and Carl Parker and Carly Burton and Catalina Mejia and Ce Liu and Changhan Wang and Changkyu Kim and Chao Zhou and Chester Hu and Ching-Hsiang Chu and Chris Cai and Chris Tindal and Christoph Feichtenhofer and Cynthia Gao and Damon Civin and Dana Beaty and Daniel Kreymer and Daniel Li and David Adkins and David Xu and Davide Testuggine and Delia David and Devi Parikh and Diana Liskovich and Didem Foss and Dingkang Wang and Duc Le and Dustin Holland and Edward Dowling and Eissa Jamil and Elaine Montgomery and Eleonora Presani and Emily Hahn and Emily Wood and Eric-Tuan Le and Erik Brinkman and Esteban Arcaute and Evan Dunbar and Evan Smothers and Fei Sun and Felix Kreuk and Feng Tian and Filippos Kokkinos and Firat Ozgenel and Francesco Caggioni and Frank Kanayet and Frank Seide and Gabriela Medina Florez and Gabriella Schwarz and Gada Badeer and Georgia Swee and Gil Halpern and Grant Herman and Grigory Sizov and Guangyi and Zhang and Guna Lakshminarayanan and Hakan Inan and Hamid Shojanazeri and Han Zou and Hannah Wang and Hanwen Zha and Haroun Habeeb and Harrison Rudolph and Helen Suk and Henry Aspegren and Hunter Goldman and Hongyuan Zhan and Ibrahim Damlaj and Igor Molybog and Igor Tufanov and Ilias Leontiadis and Irina-Elena Veliche and Itai Gat and Jake Weissman and James Geboski and James Kohli and Janice Lam and Japhet Asher and Jean-Baptiste Gaya and Jeff Marcus and Jeff Tang and Jennifer Chan and Jenny Zhen and Jeremy Reizenstein and Jeremy Teboul and Jessica Zhong and Jian Jin and Jingyi Yang and Joe Cummings and Jon Carvill and Jon Shepard and Jonathan McPhie and Jonathan Torres and Josh Ginsburg and Junjie Wang and Kai Wu and Kam Hou U and Karan Saxena and Kartikay Khandelwal and Katayoun Zand and Kathy Matosich and Kaushik Veeraraghavan and Kelly Michelena and Keqian Li and Kiran Jagadeesh and Kun Huang and Kunal Chawla and Kyle Huang and Lailin Chen and Lakshya Garg and Lavender A and Leandro Silva and Lee Bell and Lei Zhang and Liangpeng Guo and Licheng Yu and Liron Moshkovich and Luca Wehrstedt and Madian Khabsa and Manav Avalani and Manish Bhatt and Martynas Mankus and Matan Hasson and Matthew Lennie and Matthias Reso and Maxim Groshev and Maxim Naumov and Maya Lathi and Meghan Keneally and Miao Liu and Michael L. Seltzer and Michal Valko and Michelle Restrepo and Mihir Patel and Mik Vyatskov and Mikayel Samvelyan and Mike Clark and Mike Macey and Mike Wang and Miquel Jubert Hermoso and Mo Metanat and Mohammad Rastegari and Munish Bansal and Nandhini Santhanam and Natascha Parks and Natasha White and Navyata Bawa and Nayan Singhal and Nick Egebo and Nicolas Usunier and Nikhil Mehta and Nikolay Pavlovich Laptev and Ning Dong and Norman Cheng and Oleg Chernoguz and Olivia Hart and Omkar Salpekar and Ozlem Kalinli and Parkin Kent and Parth Parekh and Paul Saab and Pavan Balaji and Pedro Rittner and Philip Bontrager and Pierre Roux and Piotr Dollar and Polina Zvyagina and Prashant Ratanchandani and Pritish Yuvraj and Qian Liang and Rachad Alao and Rachel Rodriguez and Rafi Ayub and Raghotham Murthy and Raghu Nayani and Rahul Mitra and Rangaprabhu Parthasarathy and Raymond Li and Rebekkah Hogan and Robin Battey and Rocky Wang and Russ Howes and Ruty Rinott and Sachin Mehta and Sachin Siby and Sai Jayesh Bondu and Samyak Datta and Sara Chugh and Sara Hunt and Sargun Dhillon and Sasha Sidorov and Satadru Pan and Saurabh Mahajan and Saurabh Verma and Seiji Yamamoto and Sharadh Ramaswamy and Shaun Lindsay and Shaun Lindsay and Sheng Feng and Shenghao Lin and Shengxin Cindy Zha and Shishir Patil and Shiva Shankar and Shuqiang Zhang and Shuqiang Zhang and Sinong Wang and Sneha Agarwal and Soji Sajuyigbe and Soumith Chintala and Stephanie Max and Stephen Chen and Steve Kehoe and Steve Satterfield and Sudarshan Govindaprasad and Sumit Gupta and Summer Deng and Sungmin Cho and Sunny Virk and Suraj Subramanian and Sy Choudhury and Sydney Goldman and Tal Remez and Tamar Glaser and Tamara Best and Thilo Koehler and Thomas Robinson and Tianhe Li and Tianjun Zhang and Tim Matthews and Timothy Chou and Tzook Shaked and Varun Vontimitta and Victoria Ajayi and Victoria Montanez and Vijai Mohan and Vinay Satish Kumar and Vishal Mangla and Vlad Ionescu and Vlad Poenaru and Vlad Tiberiu Mihailescu and Vladimir Ivanov and Wei Li and Wenchen Wang and Wenwen Jiang and Wes Bouaziz and Will Constable and Xiaocheng Tang and Xiaojian Wu and Xiaolan Wang and Xilun Wu and Xinbo Gao and Yaniv Kleinman and Yanjun Chen and Ye Hu and Ye Jia and Ye Qi and Yenda Li and Yilin Zhang and Ying Zhang and Yossi Adi and Youngjin Nam and Yu and Wang and Yu Zhao and Yuchen Hao and Yundi Qian and Yunlu Li and Yuzi He and Zach Rait and Zachary DeVito and Zef Rosnbrick and Zhaoduo Wen and Zhenyu Yang and Zhiwei Zhao and Zhiyu Ma},
      year={2024},
      eprint={2407.21783},
      archivePrefix={arXiv},
      primaryClass={cs.AI},
      url={https://arxiv.org/abs/2407.21783}, 
}

@article{kwiatkowski-etal-2019-natural,
    title = "Natural Questions: A Benchmark for Question Answering Research",
    author = "Kwiatkowski, Tom  and
      Palomaki, Jennimaria  and
      Redfield, Olivia  and
      Collins, Michael  and
      Parikh, Ankur  and
      Alberti, Chris  and
      Epstein, Danielle  and
      Polosukhin, Illia  and
      Devlin, Jacob  and
      Lee, Kenton  and
      Toutanova, Kristina  and
      Jones, Llion  and
      Kelcey, Matthew  and
      Chang, Ming-Wei  and
      Dai, Andrew M.  and
      Uszkoreit, Jakob  and
      Le, Quoc  and
      Petrov, Slav",
    editor = "Lee, Lillian  and
      Johnson, Mark  and
      Roark, Brian  and
      Nenkova, Ani",
    journal = "Transactions of the Association for Computational Linguistics",
    volume = "7",
    year = "2019",
    address = "Cambridge, MA",
    publisher = "MIT Press",
    url = "https://aclanthology.org/Q19-1026/",
    doi = "10.1162/tacl_a_00276",
    pages = "452--466"
}

@article{trivedi2022musique,
  title={MuSiQue: Multihop Questions via Single-hop Question Composition},
  author={Trivedi, Harsh and Balasubramanian, Niranjan and Khot, Tushar and Sabharwal, Ashish},
  journal={Transactions of the Association for Computational Linguistics},
  volume={10},
  pages={539--554},
  year={2022},
  publisher={MIT Press One Broadway, 12th Floor, Cambridge, Massachusetts 02142, USA~…}
}

@inproceedings{yang2018hotpotqa,
  title={HotpotQA: A dataset for diverse, explainable multi-hop question answering},
  author={Yang, Zhilin and Qi, Peng and Zhang, Saizheng and Bengio, Yoshua and Cohen, William and Salakhutdinov, Ruslan and Manning, Christopher D},
  booktitle={Proceedings of the 2018 conference on empirical methods in natural language processing},
  pages={2369--2380},
  year={2018}
}

@inproceedings{reimers-gurevych-2019-sentence,
    title = "Sentence-{BERT}: Sentence Embeddings using {S}iamese {BERT}-Networks",
    author = "Reimers, Nils  and
      Gurevych, Iryna",
    editor = "Inui, Kentaro  and
      Jiang, Jing  and
      Ng, Vincent  and
      Wan, Xiaojun",
    booktitle = "Proceedings of the 2019 Conference on Empirical Methods in Natural Language Processing and the 9th International Joint Conference on Natural Language Processing (EMNLP-IJCNLP)",
    month = nov,
    year = "2019",
    address = "Hong Kong, China",
    publisher = "Association for Computational Linguistics",
    url = "https://aclanthology.org/D19-1410/",
    doi = "10.18653/v1/D19-1410",
    pages = "3982--3992"
}

@article{crestani1998document,
  title={“Is this document relevant?… probably” a survey of probabilistic models in information retrieval},
  author={Crestani, Fabio and Lalmas, Mounia and Van Rijsbergen, Cornelis J and Campbell, Iain},
  journal={ACM Computing Surveys (CSUR)},
  volume={30},
  number={4},
  pages={528--552},
  year={1998},
  publisher={ACM New York, NY, USA}
}

@book{gonzalez2009digital,
  title={Digital image processing},
  author={Gonzalez, Rafael C},
  year={2009},
  publisher={Pearson education india}
}

@article{marr1980theory,
  title={Theory of edge detection},
  author={Marr, David and Hildreth, Ellen},
  journal={Proceedings of the Royal Society of London. Series B. Biological Sciences},
  volume={207},
  number={1167},
  pages={187--217},
  year={1980},
  publisher={The Royal Society London}
}

@misc{openai2024textembed3,
  title        = {Text Embedding 3 Models},
  author       = {OpenAI},
  year         = {2024},
  howpublished = {\url{https://platform.openai.com/docs/models/text-embedding-3-large}},
  note         = {Accessed: 2025-12-25}
}

@inproceedings{yu-etal-2025-tree,
    title = "Tree of Agents: Improving Long-Context Capabilities of Large Language Models through Multi-Perspective Reasoning",
    author = "Yu, Song  and
      Xu, Xiaofei  and
      Deng, Ke  and
      Li, Li  and
      Tian, Lin",
    editor = "Christodoulopoulos, Christos  and
      Chakraborty, Tanmoy  and
      Rose, Carolyn  and
      Peng, Violet",
    booktitle = "Findings of the Association for Computational Linguistics: EMNLP 2025",
    month = nov,
    year = "2025",
    address = "Suzhou, China",
    publisher = "Association for Computational Linguistics",
    url = "https://aclanthology.org/2025.findings-emnlp.246/",
    doi = "10.18653/v1/2025.findings-emnlp.246",
    pages = "4574--4592",
    ISBN = "979-8-89176-335-7"
}

@article{li2023towards,
  title={Towards general text embeddings with multi-stage contrastive learning},
  author={Li, Zehan and Zhang, Xin and Zhang, Yanzhao and Long, Dingkun and Xie, Pengjun and Zhang, Meishan},
  journal={arXiv preprint arXiv:2308.03281},
  year={2023}
}

@article{estornell2024multi,
  title={Multi-LLM debate: Framework, principals, and interventions},
  author={Estornell, Andrew and Liu, Yang},
  journal={Advances in Neural Information Processing Systems},
  volume={37},
  pages={28938--28964},
  year={2024}
}

@article{qwen2.5,
    title   = {Qwen2.5 Technical Report}, 
    author  = {An Yang and Baosong Yang and Beichen Zhang and Binyuan Hui and Bo Zheng and Bowen Yu and Chengyuan Li and Dayiheng Liu and Fei Huang and Haoran Wei and Huan Lin and Jian Yang and Jianhong Tu and Jianwei Zhang and Jianxin Yang and Jiaxi Yang and Jingren Zhou and Junyang Lin and Kai Dang and Keming Lu and Keqin Bao and Kexin Yang and Le Yu and Mei Li and Mingfeng Xue and Pei Zhang and Qin Zhu and Rui Men and Runji Lin and Tianhao Li and Tingyu Xia and Xingzhang Ren and Xuancheng Ren and Yang Fan and Yang Su and Yichang Zhang and Yu Wan and Yuqiong Liu and Zeyu Cui and Zhenru Zhang and Zihan Qiu},
    journal = {arXiv preprint arXiv:2412.15115},
    year    = {2024}
}

@inproceedings{ICLR2024_5e5fd18f,
 author = {Xiao, Guangxuan and Tian, Yuandong and Chen, Beidi and Han, Song and Lewis, Mike },
 booktitle = {International Conference on Learning Representations},
 editor = {B. Kim and Y. Yue and S. Chaudhuri and K. Fragkiadaki and M. Khan and Y. Sun},
 pages = {21875--21895},
 title = {Efficient Streaming Language Models with Attention Sinks},
 url = {https://proceedings.iclr.cc/paper_files/paper/2024/file/5e5fd18f863cbe6d8ae392a93fd271c9-Paper-Conference.pdf},
 volume = {2024},
 year = {2024}
}

@inproceedings{ICLR2025_f1b04fac,
 author = {Gu, Xiangming and Pang, Tianyu and Du, Chao and Liu, Qian and Zhang, Fengzhuo and Du, Cunxiao and Wang, Ye and Lin, Min},
 booktitle = {International Conference on Learning Representations},
 editor = {Y. Yue and A. Garg and N. Peng and F. Sha and R. Yu},
 pages = {97114--97144},
 title = {When Attention Sink Emerges in Language Models: An Empirical View},
 url = {https://proceedings.iclr.cc/paper_files/paper/2025/file/f1b04face60081b689ba740d39ea8f37-Paper-Conference.pdf},
 volume = {2025},
 year = {2025}
}

@inproceedings{he-etal-2024-never,
    title = "Never Lost in the Middle: Mastering Long-Context Question Answering with Position-Agnostic Decompositional Training",
    author = "He, Junqing  and
      Pan, Kunhao  and
      Dong, Xiaoqun  and
      Song, Zhuoyang  and
      Liu, Yibo  and
      Sun, Qianguo  and
      Liang, Yuxin  and
      Wang, Hao  and
      Zhang, Enming  and
      Zhang, Jiaxing",
    editor = "Ku, Lun-Wei  and
      Martins, Andre  and
      Srikumar, Vivek",
    booktitle = "Proceedings of the 62nd Annual Meeting of the Association for Computational Linguistics (Volume 1: Long Papers)",
    month = aug,
    year = "2024",
    address = "Bangkok, Thailand",
    publisher = "Association for Computational Linguistics",
    url = "https://aclanthology.org/2024.acl-long.736/",
    doi = "10.18653/v1/2024.acl-long.736",
    pages = "13628--13642"
}

@inproceedings{han2024lm,
  title={Lm-infinite: Zero-shot extreme length generalization for large language models},
  author={Han, Chi and Wang, Qifan and Peng, Hao and Xiong, Wenhan and Chen, Yu and Ji, Heng and Wang, Sinong},
  booktitle={Proceedings of the 2024 Conference of the North American Chapter of the Association for Computational Linguistics: Human Language Technologies (Volume 1: Long Papers)},
  pages={3991--4008},
  year={2024}
}

@inproceedings{pan2024not,
  title={Not all contexts are equal: Teaching llms credibility-aware generation},
  author={Pan, Ruotong and Cao, Boxi and Lin, Hongyu and Han, Xianpei and Zheng, Jia and Wang, Sirui and Cai, Xunliang and Sun, Le},
  booktitle={Proceedings of the 2024 Conference on Empirical Methods in Natural Language Processing},
  pages={19844--19863},
  year={2024}
}

@article{wang2025jina,
  title = {jina-reranker-v3: Last but Not Late Interaction for Document Reranking},
  author={Wang, Feng and Li, Yuqing and Xiao, Han},
  journal={arXiv preprint arXiv:2509.25085},
  year={2025}
}

@article{an2024make,
  title={Make your llm fully utilize the context},
  author={An, Shengnan and Ma, Zexiong and Lin, Zeqi and Zheng, Nanning and Lou, Jian-Guang and Chen, Weizhu},
  journal={Advances in Neural Information Processing Systems},
  volume={37},
  pages={62160--62188},
  year={2024}
}

@inproceedings{hutter2025lost,
  title={Lost but not only in the middle: Positional bias in retrieval augmented generation},
  author={Hutter, Jan and Rau, David and Marx, Maarten and Kamps, Jaap},
  booktitle={European Conference on Information Retrieval},
  pages={247--261},
  year={2025},
  organization={Springer}
}

@article{yu2024mitigate,
  title={Mitigate position bias in large language models via scaling a single dimension},
  author={Yu, Yijiong and Jiang, Huiqiang and Luo, Xufang and Wu, Qianhui and Lin, Chin-Yew and Li, Dongsheng and Yang, Yuqing and Huang, Yongfeng and Qiu, Lili},
  journal={arXiv preprint arXiv:2406.02536},
  year={2024}
}

@inproceedings{wang2024leave,
  title={Leave no document behind: Benchmarking long-context llms with extended multi-doc qa},
  author={Wang, Minzheng and Chen, Longze and Cheng, Fu and Liao, Shengyi and Zhang, Xinghua and Wu, Bingli and Yu, Haiyang and Xu, Nan and Zhang, Lei and Luo, Run and others},
  booktitle={Proceedings of the 2024 Conference on Empirical Methods in Natural Language Processing},
  pages={5627--5646},
  year={2024}
}

@article{verma2024contextual,
  title={Contextual compression in retrieval-augmented generation for large language models: A survey},
  author={Verma, Sourav},
  journal={arXiv preprint arXiv:2409.13385},
  year={2024}
}

@inproceedings{du2024improving,
  title={Improving factuality and reasoning in language models through multiagent debate},
  author={Du, Yilun and Li, Shuang and Torralba, Antonio and Tenenbaum, Joshua B and Mordatch, Igor},
  booktitle={Forty-first international conference on machine learning},
  year={2024}
}

@inproceedings{chen-etal-2025-magicore,
    title = "{MA}g{IC}o{R}e: Multi-Agent, Iterative, Coarse-to-Fine Refinement for Reasoning",
    author = "Chen, Justin  and
      Prasad, Archiki  and
      Saha, Swarnadeep  and
      Stengel-Eskin, Elias  and
      Bansal, Mohit",
    editor = "Christodoulopoulos, Christos  and
      Chakraborty, Tanmoy  and
      Rose, Carolyn  and
      Peng, Violet",
    booktitle = "Proceedings of the 2025 Conference on Empirical Methods in Natural Language Processing",
    month = nov,
    year = "2025",
    address = "Suzhou, China",
    publisher = "Association for Computational Linguistics",
    url = "https://aclanthology.org/2025.emnlp-main.1660/",
    doi = "10.18653/v1/2025.emnlp-main.1660",
    pages = "32663--32686",
    ISBN = "979-8-89176-332-6"
}

@article{zhang2025qwen3,
  title={Qwen3 embedding: Advancing text embedding and reranking through foundation models},
  author={Zhang, Yanzhao and Li, Mingxin and Long, Dingkun and Zhang, Xin and Lin, Huan and Yang, Baosong and Xie, Pengjun and Yang, An and Liu, Dayiheng and Lin, Junyang and others},
  journal={arXiv preprint arXiv:2506.05176},
  year={2025}
}

@inproceedings{cormack2009reciprocal,
  author    = {Gordon V. Cormack and Charles L. A. Clarke and Stefan B{\"u}ttcher},
  title     = {Reciprocal Rank Fusion Outperforms Condorcet and Individual Rank Learning Methods},
  booktitle = {Proceedings of the 32nd International ACM SIGIR Conference on Research and Development in Information Retrieval},
  pages     = {758--759},
  year      = {2009},
  publisher = {ACM},
  doi       = {10.1145/1571941.1572114}
}

\appendix

\begin{table*}[t]
\centering
\small
\setlength{\tabcolsep}{4.5pt}
\renewcommand{\arraystretch}{1.0}

\begin{tabular*}{\textwidth}{
@{\extracolsep{\fill}}
l
l
c
l
@{}
}
\toprule
\textbf{Experiment}
& \textbf{Dataset}
& \textbf{Samples}
& \textbf{Reference} \\
\midrule

W-shape Visualization
& NQ
& $100 \times 5$
& Figure~\ref{fig:attention-wpattern} \\

\midrule
\multirow{3}{*}{Main Experiment}
& NQ
& $1{,}000 \times 5$
& \multirow{3}{*}{Table~\ref{tab:main_results}} \\
& MuSiQue
& 1,000
& \\
& HotpotQA
& 1,000
& \\

30-Document Robustness
& NQ-15
& 1,000
& Table~\ref{tab:30doc_robustness} \\

\midrule
Gold Recall Analysis
& NQ
& $100 \times 5$
& Figure~\ref{fig:fig3} \\

DocScore / Important-Doc Statistics
& NQ
& $1{,}000 \times 5$
& Table~\ref{tab:docscore_stats} \\

Cross-Model Evidence Stability
& NQ
& $1{,}000 \times 5$
& Table~\ref{tab:cross_model_docscore} \\

\midrule
\multirow{3}{*}{Component Ablation}
& NQ-15
& 1,000
& \multirow{3}{*}{Table~\ref{tab:localquality}} \\
& MuSiQue
& 1,000
& \\
& HotpotQA
& 1,000
& \\

\multirow{3}{*}{Local--Global Disagreement}
& NQ
& 1,000
& \multirow{3}{*}{Table~\ref{tab:agent_disagreement}} \\
& MuSiQue
& 1,000
& \\
& HotpotQA
& 1,000
& \\

\midrule
Local-Selection Ablation
& NQ-15
& 1,000
& Figure~\ref{fig:fig5} \\

Top-$k$ Analysis
& NQ-15
& 1,000
& Figure~\ref{fig:topk} \\

Cost Analysis
& NQ-15
& 1,000
& Table~\ref{tab:cost_comparison} \\

\bottomrule
\end{tabular*}

\caption{
Summary of experimental settings and datasets.
For NQ experiments evaluated over multiple gold positions, the sample count
is shown as (examples per position) $\times$ (number of positions);
the five positions are $n \in \{1,5,10,15,20\}$.
}
\label{tab:experiment_settings}
\end{table*}

\begin{table*}[t]
\centering
\small
\renewcommand{\arraystretch}{1.08}

\begin{tabular}{
@{}l
c
c
c
l
@{\hspace{7pt}}
c
@{\hspace{3pt}}
c
@{\hspace{3pt}}
c@{}
}
\toprule
\multirow{2}{*}{\textbf{Dataset}}
& \multirow{2}{*}{\textbf{Samples}}
& \multirow{2}{*}{\textbf{Docs}}
& \multirow{2}{*}{\textbf{Gold Docs}}
& \multirow{2}{*}{\textbf{Question Type}}
& \multicolumn{3}{c}{\textbf{Input Tokens}} \\
\cmidrule(lr){6-8}
&
&
&
&
&
\textbf{Mean}
& \textbf{Std.}
& \textbf{Max} \\
\midrule

NQ
& 1,000
& 20
& 1
& Single-hop factoid
& 2,781
& 140
& 3,396 \\

MuSiQue
& 1,000
& 20
& 2--4
& 2--4 hop
& 2,446
& 495
& 4,751 \\

HotpotQA
& 1,000
& 10
& 2
& 2-hop
& 1,351
& 349
& 3,220 \\

\midrule

NQ (30 docs)
& 1,000
& 30
& 1
& Single-hop factoid
& 4,239
& 205
& 5,474 \\

\bottomrule
\end{tabular}

\caption{
Dataset characteristics and input context statistics.
Input lengths include all documents and the question and are measured using
the Llama-3.1-8B-Instruct tokenizer.
For NQ, each positional split contains 1{,}000 examples; the 30-document
setting is an additional robustness evaluation with the gold document at position 15.
}
\label{tab:dataset_statistics}
\end{table*}

\section{Baseline Implementation Details}
\label{app:baseline}

\noindent
\emph{\textbf{(1) Vanilla LLM.}}
Constructs the input context by directly concatenating one ground-truth document with 19 distractors without additional post-processing.

\noindent
\emph{\textbf{(2) Reordering and Reranking.}}
(i) SBERT Reordering \cite{reimers-gurevych-2019-sentence} is a dense retrieval baseline using Sentence-BERT. Documents are rearranged in descending order of cosine similarity between the query and document embeddings.
(ii) BM25 Reordering \cite{crestani1998document} is a sparse retrieval baseline utilizing lexical overlap. Documents are re-ranked based on BM25 scores to prioritize exact term matches.
(iii) OpenAI Reordering \cite{openai2024textembed3} reorders documents based on semantic similarity using proprietary embeddings.
(iv) Jina-Reranker \cite{wang2025jina} is a reranking baseline based on cross-encoder-style query--document interaction. Candidate documents are re-ranked according to their relevance scores, enabling more precise ordering than bi-encoder similarity alone.
(v) Qwen3-Embedding \cite{zhang2025qwen3} is a dense embedding baseline using a recent open-source embedding model. Documents are rearranged based on cosine similarity between query and document embeddings.
(vi) Hybrid (RRF) combines the BM25 and Qwen3-Embedding rankings using Reciprocal Rank Fusion (RRF) \cite{cormack2009reciprocal}. RRF assigns each document a fused score by summing reciprocal rank-based scores across the two ranked lists:
\begin{equation}
\mathrm{RRF}(d)
=
\sum_{m \in \{\mathrm{BM25},\,\mathrm{Dense}\}}
\frac{1}{k + r_m(d)},
\end{equation}

where $r_m(d)$ denotes the rank of document $d$ under retrieval method $m$. We use $k=60$ and reorder documents according to the resulting fused scores, with Qwen3-Embedding serving as the dense retriever.
(vii) Hybrid (RRF) + Rerank further applies Jina-Reranker to the RRF-fused ranking, combining sparse and dense retrieval signals with cross-encoder reranking.

\noindent
\emph{\textbf{(3) Multi-Agent Reasoning.}}
(i) Tree of Agents (ToA) \cite{yu-etal-2025-tree} organizes agents in a hierarchical tree topology. A root agent decomposes tasks while leaf agents explore distinct paths, aggregating insights to mitigate information loss in long contexts.
(ii) MAIN-RAG \cite{chang2025main} is a collaborative architecture with specialized agents. It employs adaptive filtering to dynamically set inclusion thresholds, retaining high-confidence evidence while discarding noise.

\noindent
\emph{\textbf{(4) Context Optimization.}}
(i) Found in the Middle (FiM) \cite{hsieh2024found} is a position-aware strategy for mitigating the Lost in the Middle phenomenon. It applies a U-shaped reordering strategy that places the most relevant documents at the beginning and end of the context.
(ii) LongLLMLingua \cite{jiang2024longllmlingua} focuses on prompt compression. Following the original implementation settings, it utilizes \texttt{Llama-2-7b-chat-hf}\footnote{https://huggingface.co/meta-llama/Llama-2-7b-chat-hf} to estimate token perplexity, discarding low-importance tokens to compress the prompt while preserving key information.

\section{Experimental Settings and Dataset Details}
\label{app:dataset}

This appendix describes the datasets and experimental configurations used throughout the paper.
Table~\ref{tab:dataset_statistics} summarizes the main characteristics of each benchmark, including the number of evaluation samples, documents, gold documents, question type, and input context length statistics.










\paragraph{Natural Questions (NQ).}
The original NQ corpus \cite{kwiatkowski-etal-2019-natural} contains 307,373 training examples derived from real Google search queries, where annotators mark long answers and short answers on Wikipedia pages. In our evaluation, we use the controlled positional bias protocol of \cite{liu2024lost} rather than the original reading comprehension task. Each instance places one gold document at a designated position among 19 distractors within a 20-document sequence. Five positional splits are constructed for $n \in \{1, 5, 10, 15, 20\}$, with 1{,}000 examples per split, and all questions are single-hop. The five splits share the same document sets and differ only in the position of the gold document.

\paragraph{MuSiQue.}
MuSiQue \cite{trivedi2022musique} is constructed via bottom-up composition of single-hop questions drawn from five seed datasets into connected multi-hop chains. The dataset contains approximately 25K questions spanning 2--4 hops across six reasoning graph types. Each question enforces that one reasoning step critically depends on another, and no single paragraph suffices for an independent answer, preventing the disconnected reasoning shortcuts observed in prior benchmarks. We use instances from the answerable subset, each with its original supporting and distractor paragraphs.

\paragraph{HotpotQA.}
HotpotQA \cite{yang2018hotpotqa} contains 112,779 Wikipedia-based question--answer pairs annotated with sentence-level supporting facts. Each question requires 2-hop reasoning and is categorized as either bridge, where an intermediate entity from one paragraph is needed to locate the answer in another, or comparison, which requires contrasting properties across two paragraphs. We evaluate in the distractor setting, where two gold paragraphs are presented alongside eight distractors.

Table~\ref{tab:dataset_statistics} summarizes the benchmark-level evaluation characteristics and context length statistics, while the datasets and sample sizes used for individual experiments vary depending on the experimental purpose. All experiment-specific settings are summarized in Table~\ref{tab:experiment_settings}.

\section{DocScore Threshold Sensitivity Analysis}
\label{app:docscore_threshold}

\begin{table}[t]
\centering
\normalsize
\begin{tabular}{c|c}
\hline
\textbf{Threshold ($\tau$)} & \textbf{Recall (\%)} \\
\hline
0.2 & 83.57 \\
0.3 & 80.70 \\
\textbf{0.4} & \textbf{80.70} \\
0.5 & 79.38 \\
0.6 & 79.02 \\
0.7 & 75.06 \\
0.8 & 74.70 \\
\hline
\end{tabular}
\caption{
Sensitivity of answer-document recall to the DocScore threshold $\tau$ (operating range).
We report $\tau \in [0.2, 0.8]$ as extremely low thresholds behave similarly (near-pass-through),
while overly high thresholds enter a degenerate pruning regime.
}
\label{tab:docscore_threshold_oprange}
\end{table}

Table~\ref{tab:docscore_threshold_oprange} shows the recall of gold documents under varying DocScore thresholds.
Recall decreases monotonically as $\tau$ increases, reflecting the expected trade-off between pruning aggressiveness and evidence coverage.
Importantly, recall remains unchanged between $\tau=0.3$ and $\tau=0.4$ (80.70\%), indicating that raising the threshold within this range does not discard additional gold documents.
Beyond $\tau=0.4$, recall consistently declines, suggesting that stricter pruning begins to remove gold evidence.
We therefore set $\tau=0.4$ as the highest threshold that preserves recall while still enabling meaningful pruning of low-confidence documents.
These results are obtained on the NQ setting with 1{,}000 examples.

\begin{figure}[t]
  \centering
  \includegraphics[width=\linewidth]{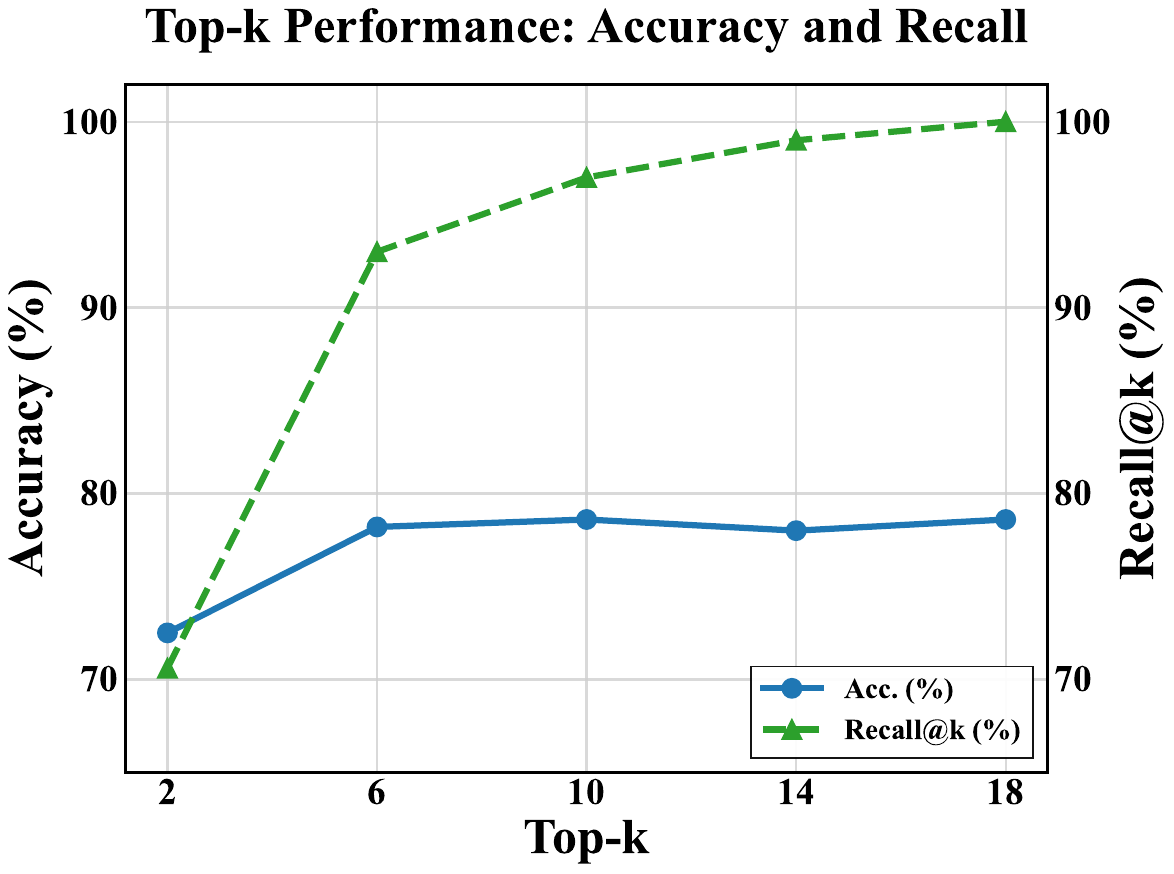}
  \caption{
  Impact of the number of documents provided to the Local Agent on Accuracy and Recall@k.
  }
  \label{fig:topk}
\end{figure}

\begin{table}[t]
\centering
\normalsize
\begin{tabular}{ccc|c}
\hline
$\Phi_{\mathrm{sim}}$ & $\Phi_{\mathrm{ppl}}$ & $\Phi_{\mathrm{cont}}$ & Avg. Recall (\%) \\
\hline
\textbf{0.5} & \textbf{1.0} & \textbf{0.5} & \textbf{91.40} \\
0.3 & 0.5 & 0.3 & 91.20 \\
0.3 & 0.7 & 0.3 & 91.20 \\
0.3 & 0.7 & 1.0 & 91.20 \\
0.3 & 0.5 & 0.7 & 91.00 \\
0.3 & 0.5 & 1.0 & 91.00 \\
0.3 & 0.7 & 0.7 & 91.00 \\
0.3 & 1.0 & 0.3 & 91.00 \\
0.5 & 0.7 & 1.0 & 91.00 \\
0.7 & 0.7 & 0.5 & 91.00 \\
\hline
\end{tabular}
\caption{Grid search results for Structural Fusion Score (SFS) weights.
Average gold evidence recall is reported across all evaluated answer positions.
The selected configuration $(\Phi_{\mathrm{sim}}, \Phi_{\mathrm{ppl}}, \Phi_{\mathrm{cont}})=(0.5,1.0,0.5)$ achieves the highest overall recall.}
\label{tab:sfs_grid}
\end{table}

\begin{table*}[t]
\centering
\small
\setlength{\tabcolsep}{3.5pt}
\renewcommand{\arraystretch}{0.92}

\begin{tabular}{@{}llcll@{}}
\hline
\textbf{Dataset (DuQUAD Acc.)}
& \textbf{Baseline}
& \textbf{Acc. (\%)}
& \textbf{Diff. (95\% CI)}
& \textbf{$p$-value} \\
\hline

\textbf{NQ-15 (76.1)}
& Vanilla LLM
& 67.0
& +9.1 (6.5 -- 11.9)
& $<$ 0.0001*** \\

& SBERT
& 70.1
& +6.0 (3.3 -- 8.8)
& $<$ 0.0001*** \\

& BM25
& 69.2
& +6.9 (4.1 -- 9.7)
& $<$ 0.0001*** \\

& Jina-Reranker
& 69.4
& +6.7 (4.1 -- 9.4)
& $<$ 0.0001*** \\

& Qwen3-Embedding
& 69.3
& +6.8 (4.0 -- 9.6)
& $<$ 0.0001*** \\

& OpenAI
& 69.9
& +6.2 (3.4 -- 9.0)
& $<$ 0.0001*** \\

& Hybrid (RRF)
& 72.2
& +3.9 (1.2 -- 6.6)
& 0.0042** \\

& \hspace{4pt}+ Rerank
& 71.4
& +4.7 (2.1 -- 7.3)
& 0.0004*** \\

& ToA
& 51.0
& +25.1 (22.0 -- 28.4)
& $<$ 0.0001*** \\

& MAIN-RAG
& 62.3
& +13.8 (10.9 -- 16.8)
& $<$ 0.0001*** \\

& LongLLMLingua
& 67.9
& +8.2 (5.5 -- 11.1)
& $<$ 0.0001*** \\

& FiM
& 64.7
& +11.4 (8.6 -- 14.2)
& $<$ 0.0001*** \\

\hline

\textbf{MuSiQue (41.1)}
& Vanilla LLM
& 33.1
& +8.0 (5.4 -- 10.6)
& $<$ 0.0001*** \\

& SBERT
& 24.9
& +16.2 (13.4 -- 18.9)
& $<$ 0.0001*** \\

& BM25
& 33.6
& +7.5 (5.1 -- 10.0)
& $<$ 0.0001*** \\

& Jina-Reranker
& 29.3
& +11.8 (9.2 -- 14.5)
& $<$ 0.0001*** \\

& Qwen3-Embedding
& 34.7
& +6.4 (3.9 -- 8.9)
& $<$ 0.0001*** \\

& OpenAI
& 34.7
& +6.4 (3.8 -- 9.1)
& $<$ 0.0001*** \\

& Hybrid (RRF)
& 28.0
& +13.1 (10.5 -- 15.8)
& $<$ 0.0001*** \\

& \hspace{4pt}+ Rerank
& 28.8
& +12.3 (9.7 -- 15.0)
& $<$ 0.0001*** \\

& ToA
& 21.5
& +19.6 (16.6 -- 22.6)
& $<$ 0.0001*** \\

& MAIN-RAG
& 19.0
& +22.1 (19.2 -- 25.0)
& $<$ 0.0001*** \\

& LongLLMLingua
& 30.8
& +10.3 (7.5 -- 13.2)
& $<$ 0.0001*** \\

& FiM
& 38.6
& +2.5 (-0.4 -- 5.4)
& 0.0910 \\

\hline

\textbf{HotpotQA (80.2)}
& Vanilla LLM
& 73.8
& +6.4 (3.9 -- 9.0)
& $<$ 0.0001*** \\

& SBERT
& 76.8
& +3.4 (1.0 -- 5.8)
& 0.0066** \\

& BM25
& 75.3
& +4.9 (2.4 -- 7.4)
& $<$ 0.0001*** \\

& Jina-Reranker
& 77.3
& +2.9 (0.4 -- 5.3)
& 0.0242* \\

& Qwen3-Embedding
& 76.8
& +3.4 (1.0 -- 5.8)
& 0.0062** \\

& OpenAI
& 76.7
& +3.5 (1.0 -- 6.0)
& 0.0072** \\

& Hybrid (RRF)
& 77.1
& +3.1 (0.5 -- 5.7)
& 0.0228* \\

& \hspace{4pt}+ Rerank
& 77.2
& +3.0 (0.5 -- 5.5)
& 0.0216* \\

& ToA
& 57.3
& +22.9 (19.8 -- 26.1)
& $<$ 0.0001*** \\

& MAIN-RAG
& 60.5
& +19.7 (16.7 -- 22.7)
& $<$ 0.0001*** \\

& LongLLMLingua
& 70.2
& +10.0 (7.2 -- 12.8)
& $<$ 0.0001*** \\

& FiM
& 76.9
& +3.3 (0.8 -- 5.9)
& 0.0106* \\

\hline
\end{tabular}

\caption{
Performance comparison of Accuracy (\%).
Values in parentheses represent the accuracy of DuQUAD.
Mean differences and 95\% confidence intervals (CI) are derived from paired
bootstrap resampling ($N=10,000$).
Asterisks indicate statistical significance:
* $p < 0.05$, ** $p < 0.01$, *** $p < 0.001$.
}
\label{tab:tab6}
\end{table*}

\section{Recall--Accuracy Trade-off for Local Document Selection}
\label{app:topk_analysis}

Figure~\ref{fig:topk} shows the effect of the local document budget $k$ on NQ-15.
Recall@$k$ increases monotonically as more documents are included and begins to
saturate around $k=6$.
Answer accuracy exhibits a similar trend, improving substantially up to $k=6$
while showing only marginal gains or slight fluctuations beyond this point.
This suggests that additional documents beyond $k=6$ mainly increase contextual
noise rather than improving final answer quality.
We therefore use $k=6$ as the default local-document budget, balancing high recall
with controlled noise.

\section{Ablation on SFS Weight Selection}
\label{app:sfs_weight}


We determine the weighting coefficients of the Structural Fusion Score (SFS)
via a grid search over $(\Phi_{\mathrm{sim}}, \Phi_{\mathrm{ppl}}, \Phi_{\mathrm{cont}})$
conducted on the Natural Questions (NQ) dataset.
Each configuration is evaluated using the average gold evidence recall
aggregated across all answer positions.
As shown in Table~\ref{tab:sfs_grid}, multiple weight combinations achieve
comparable recall values, indicating that SFS is robust to moderate variations
in weighting.
Among them, the configuration $(0.5, 1.0, 0.5)$ attains the highest overall
average recall (91.4\%) and is therefore adopted in all experiments.
This choice places greater emphasis on structural inference signals while
retaining complementary contributions from semantic relevance and attention
contrast. These results are obtained on the NQ setting with 1{,}000 examples.

\section{Reproducibility}
\label{app:reproducibility}
\begin{figure*}
  \centering
  \includegraphics[width=\linewidth]{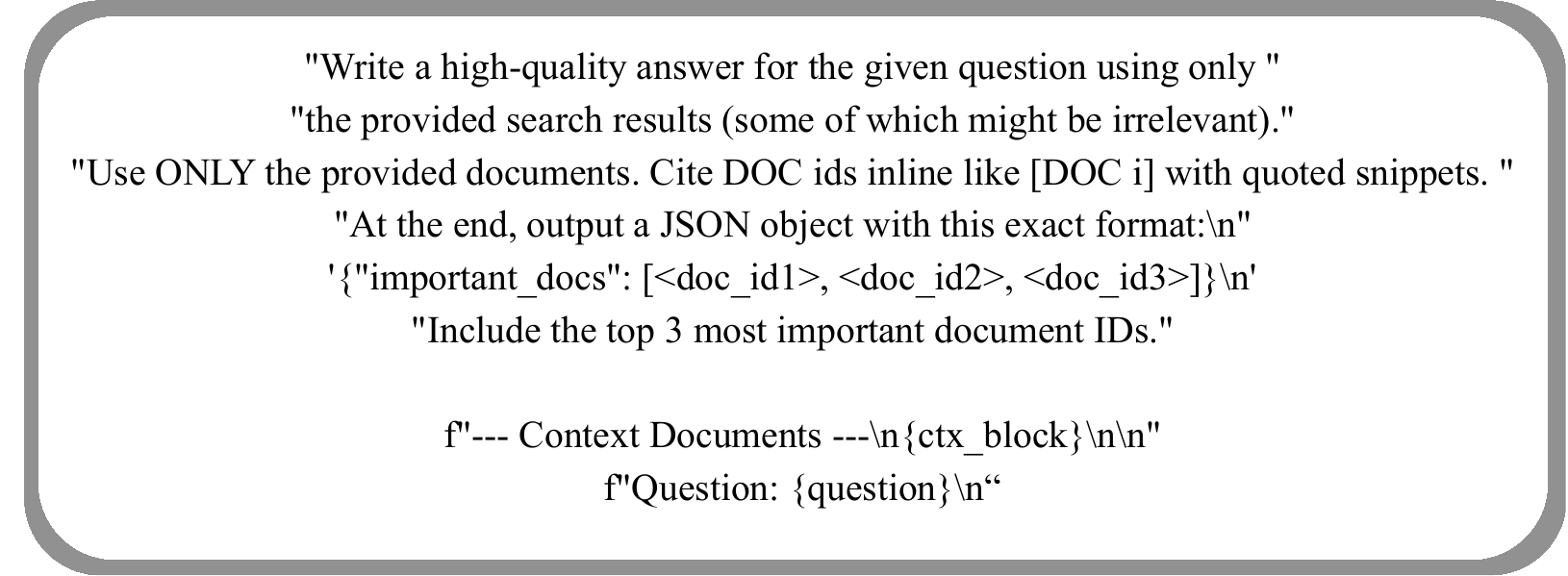}
  \caption{
  Prompt template for the Local and Global Agents.
  }
  \label{fig:prompt1}
\end{figure*}

\begin{figure*}
  \centering
  \includegraphics[width=\linewidth]{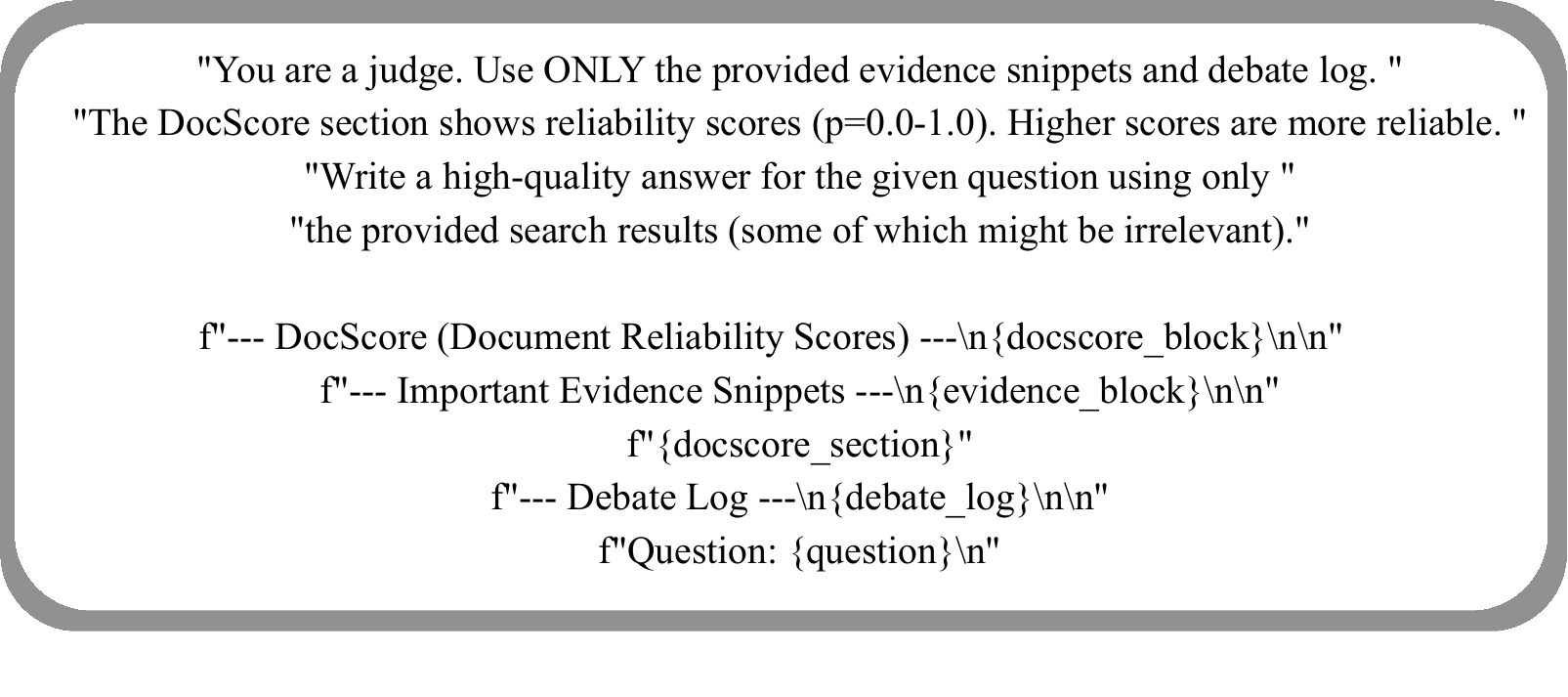}
  \caption{
  Prompt template for the Judge Agent.
  }
  \label{fig:prompt2}
\end{figure*}

\begin{figure*}
  \centering
  \includegraphics[width=\linewidth]{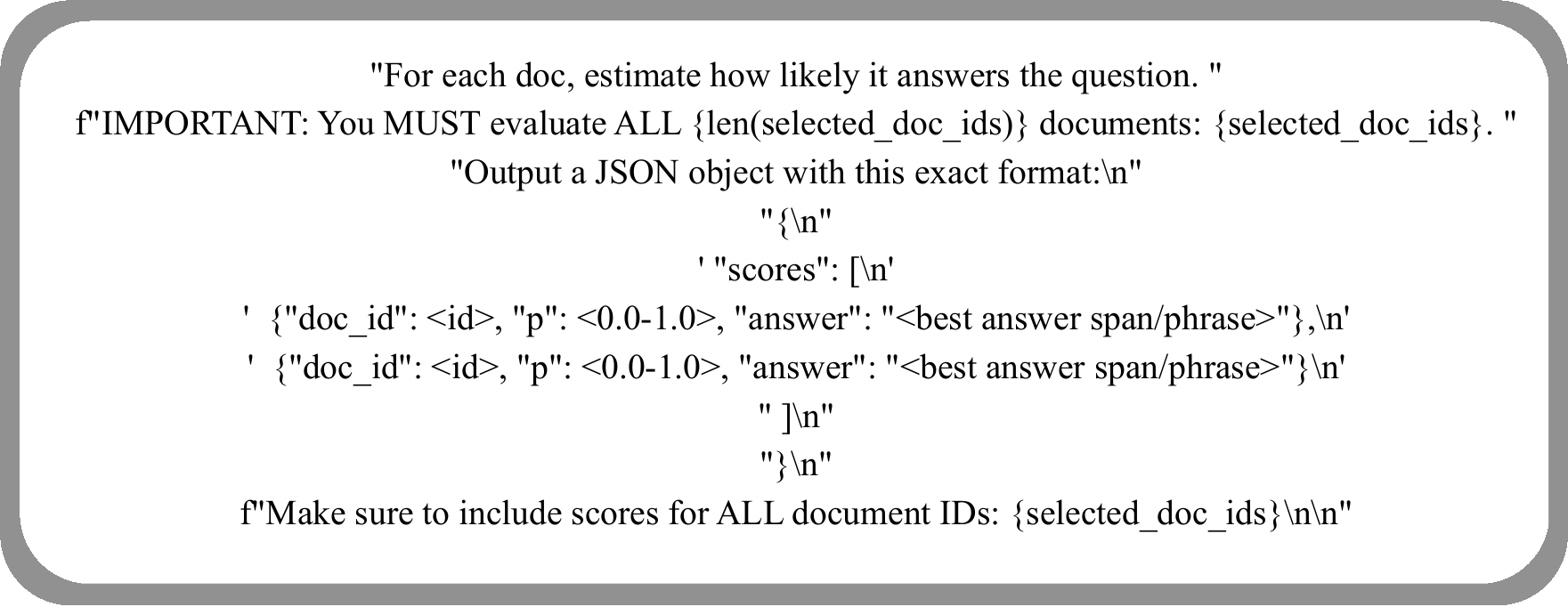}
  \caption{
  Prompt template for the Score Agent.
  }
  \label{fig:prompt3}
\end{figure*}

To ensure reproducibility, we release the exact prompts used for all LLM-based components.
Reasoning agent prompts, Judge prompts, and score estimation prompts are provided in Figures~\ref{fig:prompt1},~\ref{fig:prompt2}, and~\ref{fig:prompt3}, respectively, without modification.
In particular, the reasoning agent prompt in Figure~\ref{fig:prompt1} instructs each agent to answer the question using only the provided documents and to return the top 3 important document IDs in a fixed JSON format, which is later used to construct the candidate document pool.
The score estimation prompt in Figure~\ref{fig:prompt3} defines the DocScore procedure: for each selected candidate document, the Score Agent estimates whether the question can be answered from that document alone and returns a JSON object containing a confidence score $p \in [0,1]$ together with the best answer span or phrase.

All experiments were conducted on a computing system equipped with NVIDIA L40S GPUs, an AMD EPYC 9654 CPU (96 cores), and 1~TB of system memory.
We fix all decoding parameters across experiments and use deterministic decoding with a temperature of 0.0 unless otherwise specified, enabling faithful replication of our results.
All baseline experiments were conducted under the same generation setting with the temperature fixed to 0.0 to ensure reproducibility and fair comparison.
The basic prompt templates (e.g., Vanilla LLM and reordering-based methods) strictly follow the configurations used in the Lost in the Middle paper \footnote{https://github.com/nelson-liu/lost-in-the-middle}.
For baselines requiring additional prompt control or specialized templates, we prioritized the officially released GitHub implementations provided by the original authors.
When an official implementation was not available, we faithfully reimplemented the prompts and inference procedures as described in the corresponding papers.

\paragraph{Artifact use.}
All existing artifacts used in this work are used solely for research and evaluation purposes, consistent with their intended use.

Structural inference and attention score computation are performed using the same backbone model as the main reasoning process.
This avoids introducing separate scoring models and helps mitigate bias arising from reliance on specific model architectures.

\section{Additional Evaluation Analyses}

\subsection{Statistical Significance Analysis}
\label{statistic}

We evaluate the statistical significance of the performance differences between DuQUAD and each baseline using paired bootstrap resampling.
For each dataset, we repeatedly resample the evaluation set with replacement (10,000 iterations) and compute the containment rate difference between DuQUAD and the corresponding baseline.
The reported difference corresponds to the mean improvement, and the 95\% confidence interval (CI) is obtained from the 2.5th and 97.5th percentiles of the bootstrap distribution.
Statistical significance is assessed using a two-sided bootstrap hypothesis test under the null hypothesis that the expected difference is zero.
Two-sided p-values are computed as twice the smaller empirical tail probability of the bootstrap distribution relative to zero.
We report significance levels using standard conventions: *** for $p < 0.001$, ** for $p < 0.01$, and * for $p < 0.05$.

The complete statistical results are summarized in Table~\ref{tab:tab6}.
DuQUAD yields positive mean improvements over all baselines across the 36 pairwise comparisons (12 baselines $\times$ 3 datasets), with 35 of them reaching statistical significance at $p < 0.05$.
Significant improvements are observed across reordering and reranking methods, multi-agent frameworks, and context optimization techniques, including strong baselines such as Hybrid (RRF), Hybrid (RRF) + Rerank, Jina-Reranker, Qwen3-Embedding, LongLLMLingua, and FiM on NQ-15 and HotpotQA.

The single exception is DuQUAD vs.\ FiM on MuSiQue.
Two factors underlie this result: FiM is a strong baseline on multi-hop reasoning, as its U-shaped reordering places relevant documents at salient positions well suited to compositional evidence, and the absolute gap ($+2.5\%$) is the smallest among all MuSiQue comparisons, where the remaining gaps ($+6.4\%$ to $+22.1\%$) yield clearer statistical separation.
The direction of improvement nonetheless remains consistent, and DuQUAD shows clearer statistical advantages in settings with stronger positional bias, such as NQ.

\subsection{Robustness to Increased Document Count}
\label{app:30doc_robustness}

We further evaluate robustness to an increased number of documents by
extending NQ-15 from 20 to 30 documents while keeping the gold document
at position 15.
Table~\ref{tab:30doc_robustness} compares accuracy with the corresponding
20-document NQ-15 setting.

\begin{table}[t]
\centering
\small
\setlength{\tabcolsep}{4pt}
\renewcommand{\arraystretch}{1.08}
\begin{tabular}{lccc}
\toprule
\textbf{Method}
& \textbf{20 Docs}
& \textbf{30 Docs}
& \textbf{Diff.} \\
\midrule

Vanilla LLM
& 67.0
& 60.8
& $-6.2$ \\

OpenAI
& 69.9
& 66.1
& $-3.8$ \\

Hybrid (RRF)
& 72.2
& 64.8
& $-7.4$ \\

Hybrid (RRF) + Rerank
& 71.4
& 67.8
& $-3.6$ \\

FiM
& 64.7
& 62.7
& $-2.0$ \\

\textbf{DuQUAD}
& \textbf{76.1}
& \textbf{75.6}
& \textbf{$-0.5$} \\

\bottomrule
\end{tabular}

\caption{
Accuracy on NQ-15 under the 20- and 30-document settings.
Diff. denotes the accuracy change when increasing the number of documents
from 20 to 30.
}
\label{tab:30doc_robustness}
\end{table}

\begin{table*}[t]
\centering
\begin{tabular}{lcccccc}
\toprule
Method & \multicolumn{2}{c}{LLM Calls} & \multicolumn{2}{c}{Tokens} & \multicolumn{2}{c}{E2E Latency (s)} \\
 & Value & Ratio & Value & Ratio & Value & Ratio \\
\midrule
Vanilla LLM
& 1.00 & 1.00$\times$
& 2863.50 & 1.00$\times$
& 2.12 & 1.00$\times$ \\

OpenAI
& 1.00 & 1.00$\times$
& 2900.31 & 1.01$\times$
& 4.38 & 2.07$\times$ \\

Qwen3-Embedding
& 1.00 & 1.00$\times$
& 2940.39 & 1.03$\times$
& 3.70 & 1.75$\times$ \\

Jina-Reranker
& 1.00 & 1.00$\times$
& 2940.53 & 1.03$\times$
& 3.98 & 1.88$\times$ \\

Hybrid (RRF)
& 1.00 & 1.00$\times$
& 2939.80 & 1.03$\times$
& 3.67 & 1.73$\times$ \\

\hspace{4pt}+ Rerank
& 1.00 & 1.00$\times$
& 2934.68 & 1.02$\times$
& 3.65 & 1.72$\times$ \\

FiM
& 2.00 & 2.00$\times$
& 5437.09 & 1.90$\times$
& 4.33 & 2.04$\times$ \\

LongLLMLingua
& 2.00 & 2.00$\times$
& 5822.48 & 2.03$\times$
& 5.81 & 2.74$\times$ \\

MAIN-RAG
& 41.00 & 41.00$\times$
& 13880.14 & 4.85$\times$
& 69.22 & 32.67$\times$ \\

ToA
& 23.17 & 23.17$\times$
& 16613.13 & 5.80$\times$
& 81.58 & 38.50$\times$ \\

DuQUAD
& 4.10 & 4.10$\times$
& 5914.07 & 2.07$\times$
& 21.03 & 9.93$\times$ \\
\bottomrule
\end{tabular}
\caption{
Cost comparison normalized to Vanilla LLM in terms of LLM calls,
tokens, and end-to-end latency.
}
\label{tab:cost_comparison}
\end{table*}

As shown in Table~\ref{tab:30doc_robustness}, increasing the number of
documents substantially degrades Vanilla LLM and several competing baselines,
whereas DuQUAD remains stable, decreasing by only 0.5 points.
Among the evaluated baselines in the 30-document setting,
Hybrid (RRF) + Rerank achieves 67.8\% accuracy, leaving a 7.8-point gap
to DuQUAD's 75.6\%.
FiM also exhibits a relatively small degradation of 2.0 points, but its
absolute accuracy remains substantially lower at 62.7\%.

\section{Cost Analysis}
\label{app:cost}

Table~\ref{tab:cost_comparison} reports the inference cost of DuQUAD and representative baselines in terms of LLM calls, total token consumption, and end-to-end latency, all measured on the same 1{,}000-instance NQ-15 evaluation set used for the NQ-15 accuracy results in Table~\ref{tab:main_results} and normalized to the Vanilla setting.
LLM calls reflect the number of model invocations in each pipeline, token counts capture the effective computation incurred when processing multi-document contexts, and end-to-end latency measures the wall-clock time required to produce a final answer.
For retrieval and reranking methods, including OpenAI, Qwen3-Embedding, Jina-Reranker, and the hybrid variants, retrieval-side computations are performed by separate models rather than the answer-generation LLM, and their associated calls and tokens are therefore not included in the reported LLM call and token counts, while their computational overhead is reflected in end-to-end latency.

Table~\ref{tab:cost_comparison} shows that lightweight retrieval and reranking methods require only a single answer-generation LLM call, whereas existing multi-agent frameworks require substantially more calls and latency.
Among the lightweight methods, Hybrid (RRF) achieves the highest NQ-15 accuracy at 72.2\% with an end-to-end latency of 3.67\,s, while Qwen3-Embedding and Hybrid (RRF) + Rerank show comparable latency at 3.70\,s and 3.65\,s, respectively.
DuQUAD further improves accuracy to 76.1\%, while using $4.10$ LLM calls and incurring $9.93\times$ end-to-end latency relative to Vanilla.

Compared to MAIN-RAG, DuQUAD reduces LLM calls by $10\times$ ($4.10$ vs.\ $41.00$), token consumption by $2.35\times$ ($5{,}914$ vs.\ $13{,}880$), and end-to-end latency by $3.29\times$ ($21.03$s vs.\ $69.22$s).
Compared to ToA, DuQUAD reduces LLM calls by $5.65\times$, token consumption by $2.81\times$, and end-to-end latency by $3.88\times$.
At the same time, DuQUAD outperforms both methods in accuracy across all three benchmarks (Table~\ref{tab:main_results}), indicating that the accuracy benefits of multi-agent reasoning can be retained while substantially reducing its characteristic cost overhead.

Although DuQUAD is more costly than lightweight retrieval and reranking methods due to dual-view reasoning and quality-aware pruning, its overhead remains moderate relative to existing multi-agent frameworks and is justified when accuracy is prioritized under positional bias and noise.
\begin{table*}[t]
\centering
\small
\setlength{\tabcolsep}{3.5pt}
\renewcommand{\arraystretch}{1.05}

\begin{tabular}{@{}lcccccc@{}}
\toprule

\multirow{2}{*}{\textbf{Dataset}}
& \multirow{2}{*}{\textbf{Both Correct}}
& \multirow{2}{*}{\textbf{Global Only}}
& \multirow{2}{*}{\textbf{Local Only}}
& \multirow{2}{*}{\textbf{Both Wrong}}
& \textbf{Disagreement}
& \textbf{Judge Acc.} \\

&
&
&
&
&
\textbf{Rate}
& \textbf{on Disagreement} \\

\midrule

NQ
& 579
& 131
& 112
& 178
& 24.3\%
& 74.5\% (181/243) \\

HotpotQA
& 746
& 72
& 63
& 119
& 13.5\%
& 58.5\% (79/135) \\

MuSiQue
& 291
& 103
& 83
& 523
& 18.6\%
& 67.2\% (125/186) \\

\bottomrule
\end{tabular}

\caption{
Local--Global agent agreement and disagreement statistics.
Each dataset contains 1{,}000 evaluation examples.
Disagreement denotes cases in which exactly one agent is correct,
and Judge accuracy is measured only over these disagreement cases.
}
\label{tab:agent_disagreement}
\end{table*}

\section{Boundary Handling for Attention Contrast Computation}
\label{app:att_contrast}
When computing attention contrast, special handling is required at document boundary positions where symmetric neighbors do not exist.
Instead of constructing a second-order central difference operator that assumes symmetric neighbors through padding or reflection, we adopt a one-sided finite difference scheme at the document boundaries.
Specifically, for the first and last documents, attention contrast is computed using the difference between the document and its nearest neighbor.
This design reduces artificial smoothing effects introduced by padding-based boundary conditions while more faithfully preserving relative attention salience at the document edges.
The one-sided difference scheme is widely used across numerical analysis and signal processing, and in our experiments, it exhibits consistent and interpretable behavior for document-level contrast estimation.

\section{Document-Level Attention Scoring}
\label{app:doc_attention}

This section describes how we compute document-level attention scores, which are
used as one component of the Structural Fusion Score.

\paragraph{Input sequence construction.}
Given a question $Q$ and retrieved documents $\{D_1,\dots,D_N\}$, we follow the
query-aware contextualization format of \citet{liu2024lost}, which places the
query before and after the document sequence:
\[
[Q; D_1; D_2; \dots; D_N; Q].
\]
The second occurrence of $Q$ serves as the final query segment. In decoder-only
LLMs, the last token of this final query segment can causally attend to all
preceding document tokens. Each document $D_i$ is preceded by a delimiter token
(e.g., \texttt{[DOC]}), and its token span is recorded as
$\mathrm{span}(D_i)=[s_i,e_i)$.

\paragraph{Document-level attention score.}
Let $t^\star$ denote the position of the last token in the final query segment.
We run the backbone LLM with attention outputs enabled and extract the
self-attention vector from $t^\star$ to all previous input tokens. For each
layer $\ell$ and head $h$, this vector is denoted by
$A^{(\ell,h)}_{t^\star,:}$. We average it over all layers and heads:
\[
r[j] =
\frac{1}{LH}
\sum_{\ell=1}^{L}
\sum_{h=1}^{H}
A^{(\ell,h)}_{t^\star,j}.
\]
Here, $r[j]$ denotes the average attention mass assigned from the final query
token to token position $j$. The document-level attention score for $D_i$ is then
computed as the mean attention mass over its token span:
\[
\mathrm{AttnScore}(D_i) =
\frac{1}{|\mathrm{span}(D_i)|}
\sum_{j \in \mathrm{span}(D_i)} r[j].
\]
This computation requires access to the model's self-attention tensors, such as
through \texttt{output\_attentions=True} in open-source implementations. It does
not modify model parameters or decoding behavior, but assumes that attention
weights are available.

\section{Analysis of Local--Global Complementarity}
\label{app:failure}

To quantify the complementary behavior of the Local and Global Agents,
we categorize each prediction according to whether each agent produces the
correct answer. Table~\ref{tab:agent_disagreement} reports the resulting
statistics over 1{,}000 examples from each dataset.
We define a disagreement case as one in which exactly one of the two agents
is correct, and report Judge accuracy as the proportion of these cases for
which the final Judge Agent produces the correct answer.











Table~\ref{tab:agent_disagreement} reveals that both Global-only and
Local-only successes occur consistently across all three datasets,
indicating that neither view subsumes the other.
The disagreement rate ranges from 13.5\% on HotpotQA to 24.3\% on NQ,
and the Judge Agent resolves these disagreements correctly more often than
not on all three datasets, reaching 74.5\% accuracy on NQ, 58.5\% on
HotpotQA, and 67.2\% on MuSiQue.
These results quantitatively support the complementary roles of the two
reasoning views and the benefit of integrating them at the final decision stage.

The remaining errors also reveal a distinct limitation on more complex
multi-hop reasoning.
On MuSiQue, both agents are incorrect on 523 of 1{,}000 examples (52.3\%),
substantially more often than on NQ or HotpotQA.
This suggests that many MuSiQue failures arise before final adjudication,
when both reasoning views fail to correctly track or integrate intermediate
entities across multiple hops.

To complement this aggregate analysis, we present three representative cases
that illustrate how the Local and Global Agents interact: two cases in which
one agent fails while the other recovers the correct answer, and one failure
case in which both agents converge on the same incorrect reasoning.

\begin{table}[t]
\centering
\small
\setlength{\tabcolsep}{6pt}
\renewcommand{\arraystretch}{1.15}
\begin{tabular}{llcc}
\toprule
\textbf{Model} & \textbf{Document Type} & \textbf{DS Mean} & \textbf{ID Avg.} \\
\midrule
\multirow{2}{*}{Llama}
& Gold Doc & \textbf{0.647} & \textbf{85.7} \\
& Non-Gold Doc & 0.324 & 14.3 \\
\midrule
\multirow{2}{*}{Qwen}
& Gold Doc & \textbf{0.722} & \textbf{83.0} \\
& Non-Gold Doc & 0.348 & 17.0 \\
\bottomrule
\end{tabular}
\caption{
DocScore (DS) and Important Document (ID) statistics on Natural Questions (NQ-1/5/10/15/20).
DS Mean denotes the average DocScore assigned to documents of each type,
while ID Avg.\ indicates the probability that a document of the corresponding type is selected as an Important Document.
}
\label{tab:docscore_stats}
\end{table}

\paragraph{Case 1: Global Recovery under Local Error.}
For the question \textit{``who won the mens figure skating in 2018 olympics''}, the gold document is DOC 14, and the correct answer is Yuzuru Hanyu. The Global Agent correctly identifies the answer from the full document set by citing DOC 14. In contrast, the Local Agent predicts Nathan Chen, even though DOC 14 is included in its selected local subset. Instead of relying on the answer-bearing statement, the Local branch is distracted by another document describing Nathan Chen's technical achievements, which are salient but not sufficient to answer the question. During quality-aware pruning, DOC 14 receives a DocScore of 1.0, while the other candidate documents receive 0.0. The Judge Agent therefore rejects the Local Agent's incorrect claim and produces the correct final answer, Yuzuru Hanyu. This case shows that even when the Local branch fails, the Global branch and subsequent pruning can prevent the error from propagating to the final decision.
\paragraph{Case 2: Local Recovery under Global Uncertainty.}
We also observe cases in which the Global branch follows a more indirect reasoning path, whereas the Local branch recovers the answer more directly. For the question \textit{``who performed the halftime show at super bowl 51''}, the Global Agent initially states that the performer is unclear and refers to surrounding information that Adele had been discussed as a possible candidate. Although it eventually reaches the correct answer by citing DOC 14, its reasoning does not immediately converge on the core answer-bearing evidence. In contrast, the Local Agent explicitly states that \textit{``Lady Gaga performed the halftime show at Super Bowl LI (which is Super Bowl 51) [DOC 14],''} directly identifying the correct entity and clarifying the equivalence between Super Bowl LI and Super Bowl 51. At the final integration stage, this complementary behavior allows the system to converge stably to the correct answer, Lady Gaga. This case shows that even when one branch becomes temporarily unstable, the other branch can preserve the correct answer signal and support a stable final prediction.
\paragraph{Case 3: Joint Failure under Multi-hop Entity Tracking.}
We also observe failure cases in which DuQUAD retrieves several relevant documents but follows the wrong entity branch during final synthesis. In a MuSiQue example, the question requires a four-hop chain from Lion Air to Juanda International Airport, then to Indonesia, then to East Timor through the Indonesia--Timor Leste Commission of Truth and Friendship, and finally to the president of East Timor (Francisco Guterres). The Local Agent selects multiple gold documents, including DOC 9, which explicitly states \textit{``President Francisco Guterres,''} and quality-aware pruning assigns high DocScores to the East Timor and commission documents (DOC 9: $p = 0.80$; DOC 14: $p = 0.90$). However, both the Local and Global Agents focus on the Indonesia branch of the reasoning chain and infer the Indonesian president, Susilo Bambang Yudhoyono. The Judge Agent then follows this shared but incorrect intermediate entity instead of resolving the target country as East Timor, and outputs the wrong final answer. This case shows that although DuQUAD reliably surfaces and scores answer-bearing evidence, complex multi-hop questions can still fail when the final synthesis stage must track closely related entities across several intermediate hops.


\begin{table}[t]
\centering
\setlength{\tabcolsep}{8pt}
\renewcommand{\arraystretch}{1.15}
\begin{tabular}{lc}
\toprule
\textbf{Metric} & \textbf{Value} \\
\midrule
Gold evidence stability & 90.4\% \\
Distractor variability & 90.1\% \\
\bottomrule
\end{tabular}
\caption{
Cross-model evidence selection stability on Natural Questions (NQ-1/5/10/15/20).
}
\label{tab:cross_model_docscore}
\end{table}

\section{DocScore Statistics and Cross-Model Evidence Stability}
\label{app:cross_model}

Table~\ref{tab:docscore_stats} summarizes the DocScore and Important Document statistics used in the main analysis.
For both backbones, gold documents receive substantially higher DocScores than non-gold documents and are selected as Important Documents much more frequently.
Under Llama, gold and non-gold documents receive mean DocScores of 0.647 and 0.324, respectively, with Important Document selection rates of 85.7\% and 14.3\%.
Under Qwen, the same pattern holds: gold and non-gold documents receive mean DocScores of 0.722 and 0.348, with selection rates of 83.0\% and 17.0\%, respectively.
This consistent separation across backbones suggests that DocScore captures document-level answerability rather than simply reflecting positional artifacts or model-specific scoring behavior.

Since DocScore is produced by an LLM-based Score Agent, its evidence selection could in principle reflect backbone-specific behavior rather than an intrinsic notion of evidence quality.
To test this, we compare DocScore-based evidence selection across Llama-3.1-8B-Instruct and Qwen2.5-7B-Instruct under identical settings on Natural Questions, averaged over NQ-1/5/10/15/20.
The cross-model selection results are reported in Table~\ref{tab:cross_model_docscore}.

Our goal here is to measure cross-model \emph{agreement}, not the absolute ability of either backbone to recover the gold document.
Accordingly, when measuring gold evidence stability, we restrict the denominator to cases where at least one backbone selects the gold document, and report how often both do.
Cases where neither backbone selects the gold document are excluded, as they reflect retrieval difficulty rather than disagreement between backbones, and would otherwise conflate selection quality with cross-model consistency.

\paragraph{Disagreement is concentrated on distractors.}
The selected document sets are not identical across backbones, but the disagreement is concentrated on distractors rather than gold evidence.
When at least one backbone selects the gold document, both select it in 90.4\% of cases.
In contrast, among cases where both backbones select the gold document, their additionally selected non-gold documents differ 90.1\% of the time.

\paragraph{Implication.}
High agreement on gold evidence alongside divergence on distractors shows that cross-model consistency is not an artifact of identical candidate pools: the two backbones converge on gold evidence while diverging on irrelevant context.
DocScore therefore captures a stable answerability signal across the two evaluated backbones, supporting its use as a reliable quality signal for evidence pruning.

\begin{algorithm*}[t]
\caption{DuQUAD Pipeline for a Single Question}
\label{alg:duquad}
\textbf{Input:} question $Q$, documents $\{\mathcal{D}_i\}_{i=1}^{N}$, local top-$k$, threshold $\tau$ \\
\textbf{Output:} final answer $\hat{y}$ \\
\vspace{0.4em}
~\\
\textbf{(1) Local selection via Structural Fusion Score (SFS)} \\
$a_i \gets$ aggregated attention score of $\mathcal{D}_i$ \\
$\Phi_{\mathrm{sim},i} \gets$ semantic relevance score of $\mathcal{D}_i$ \\
$\Phi_{\mathrm{ppl},i} \gets$ structural inference score of $\mathcal{D}_i$ \\
\vspace{0.8em}
$z_i \gets \dfrac{a_i - \mu_A}{\sigma_A}$
\hfill $\triangleright$ z-score normalization
($\mu_A$, $\sigma_A$: mean/std of $\{a_i\}_{i=1}^{N}$) \\
\vspace{0.8em}
$\Phi_{\mathrm{cont},i} \gets z_i - \dfrac{z_{i-1}+z_{i+1}}{2}$
\hfill $\triangleright$ attention contrast ($i=2,\dots,N-1$) \\
\vspace{0.8em}
$\Phi_i \gets
\alpha\,\mathrm{softmax}(\Phi_{\mathrm{sim}})_i
+ \beta\,\mathrm{softmax}(-\Phi_{\mathrm{ppl}})_i
+ \gamma\,\mathrm{softmax}(\Phi_{\mathrm{cont}})_i$
\hfill $\triangleright$ structural fusion score \\
\vspace{0.8em}
$\mathcal{I}_{\mathrm{local}} \gets \operatorname*{arg\,top}_{k} \Phi_i$
\hfill $\triangleright$ local document selection \\
$\mathcal{D}_{\mathrm{local}} \gets \{\mathcal{D}_i \mid i \in \mathcal{I}_{\mathrm{local}}\}$ \\
\vspace{0.8em}
~\\
\textbf{(2) Dual-view Agent Collaboration} \\
$(\mathcal{H}_G, \mathcal{D}_G) \gets \mathrm{GlobalAgent}(Q, \{\mathcal{D}_i\}_{i=1}^{N})$
\hfill $\triangleright$ global reasoning + important docs \\
$(\mathcal{H}_L, \mathcal{D}_L) \gets \mathrm{LocalAgent}(Q, \mathcal{D}_{\mathrm{local}})$
\hfill $\triangleright$ local reasoning + important docs \\
$\mathcal{D}_{\mathrm{imp}} \gets \mathcal{D}_G \cup \mathcal{D}_L$
\hfill $\triangleright$ merge important docs \\
\vspace{0.8em}
~\\
\textbf{(3) DocScore-based Quality Gating} \\
$(p_i, \tilde{a}_i) \gets \mathrm{ScoreAgent}(Q, \mathcal{D}_i),
\quad \forall \mathcal{D}_i \in \mathcal{D}_{\mathrm{imp}}$
\hfill $\triangleright$ document reliability + answer span \\
$\mathcal{D}_{\mathrm{keep}} \gets \{\mathcal{D}_i \in \mathcal{D}_{\mathrm{imp}} \mid p_i \ge \tau\}$
\hfill $\triangleright$ DocScore-based pruning \\
\vspace{0.8em}
~\\
\textbf{(4) Evidence Snippet Extraction} \\
$\mathcal{S}_i \gets \mathrm{Top2}_j\, \mathrm{overlap}(u_{i,j}, \tilde{a}_i),
\quad \forall \mathcal{D}_i \in \mathcal{D}_{\mathrm{keep}}$
\hfill $\triangleright$ select evidence sentences \\
$\mathcal{S} \gets \bigcup_{\mathcal{D}_i \in \mathcal{D}_{\mathrm{keep}}} \mathcal{S}_i$
\hfill $\triangleright$ aggregate evidence \\
\vspace{0.8em}
~\\
\textbf{(5) Final Decision} \\
$\hat{y} \gets
\mathrm{JudgeAgent}(Q, \mathcal{H}_G, \mathcal{H}_L, \mathcal{S},
\{p_i\}_{\mathcal{D}_i \in \mathcal{D}_{\mathrm{keep}}})$
\hfill $\triangleright$ final answer generation \\
\textbf{return} $\hat{y}$
\end{algorithm*}

\end{document}